\documentclass[12pt]{article}
\usepackage{amsmath,amssymb}
\numberwithin{equation}{section}
\newcommand{\bq}{\begin{eqnarray}}
\newcommand{\eq}{\end{eqnarray}}
\newcommand{\be}{\begin{equation}}
\newcommand{\ee}{\end{equation}}
\newcommand{\ra}{\rightarrow}
\newcommand{\gd}{ \gamma_q}
\newcommand{\ov}{\overline}
\newcommand{\nd}{{\ov N}_c}
\newcommand{\la}{ \Lambda_Q }
\newcommand{\p}{{\ov Q}_{\ov j}Q^i}

\newcommand{\te}{d^2\theta}
\newcommand{\ote}{d^2\ov \theta}

\newcommand{\cm}{{\cal M}_{\rm ch}}
\newcommand{\co}{{\cal M}_{o}}
\newcommand{\lym}{\Lambda_{YM}}
\newcommand{\bo}{\rm{b_o}}
\newcommand{\bd}{\rm {\ov b}_o}

\begin{document}
\begin{center}
{\bf On Mass Spectrum in SQCD,\\ and Problems with the Seiberg Duality.\\
\, Equal quark masses}
\end{center}
\vspace{1cm}
\begin{center}\bf Victor L. Chernyak \end{center}
\begin{center}(e-mail: v.l.chernyak@inp.nsk.su) \end{center}
\begin{center} Budker Institute of Nuclear Physics, 630090 Novosibirsk-90, Russia
\end{center}
\vspace{1cm}
\begin{center}Abstract \end{center}
\vspace{1cm}
The dynamical scenario is considered for ${\cal N}=1$ SQCD, with $N_c$ colors and $N_c<N_F<3N_
c$ flavors with small but nonzero current quark masses $m_Q\neq 0$, in which quarks form the
diquark-condensate phase. This means that colorless chiral quark pairs condense coherently
in the vacuum, \,$\langle{\ov Q}Q\rangle\neq 0$, while quarks alone don't condense, $\langle Q
\rangle=\langle{\ov Q}\rangle=0$, so that the color is confined. Such condensation of
quarks results in formation of dynamical constituent masses $\mu_C\gg m_Q$ of quarks and
appearance of light "pions" (similarly to QCD). The mass spectrum
of SQCD in this phase is described and comparison with the Seiberg dual description is
performed. It is shown that the direct and dual theories are different (except, possibly, for
the perturbative strictly superconformal  regime).   \\
\vspace{1cm}

\section {Introduction}

\hspace {6 mm} Because supersymmetric gauge theories are much more constrained in comparison with
ordinary ones, it is easier for theory to deal with them. So, they can serve, at least,
as useful models for elucidating the complicated strong coupling gauge dynamics (not even
speaking about their potential relevance to a real world).

The closest to QCD is its supersymmetric extension ${\cal N}=1$ SQCD, and it was considered in
many papers. We will be dealing here with SQCD in the non-perturbative region (or in the
perturbative strong coupling regime). Most impressive results here were obtained by
N. Seiberg, who proposed description of this strongly coupled (and/or non-perturbative) SQCD
through the equivalent, but weakly coupled dual theory \cite{S1} (for reviews, see [2,3,4].

Our purpose in this paper is to introduce in section 3 the main dynamical assumption about the
coherent diquark-condensate (DC) phase of SQCD, to describe its consequences for the behavior in the
infrared region, the mass spectrum, etc., and to compare with predictions of the Seiberg
dual theory.

The paper is organized as follows. Sections 2 and 4 recall definitions of the direct and dual
theories, and some particular examples are considered in section 2.  Both direct and dual theories
are considered in the conformal window $3N_c/2 <N_F<3N_c$ in sections 3 and 5-6, respectively,
and at $N_c<N_F<3N_c/2$ in section 7. For completeness, the case $N_F> 3N_c$ is considered in
section 8. Finally, some conclusions are presented in section 9 (and there is one appendix about
't Hooft triangles).

\section { Direct theory\,. Definition and some examples.}

\hspace {6 mm} The fundamental Lagrangian of SQCD with $N_c$ colors and $N_F$ flavors (at high scale
$\mu \gg \la$) is given by:
\bq
L=\int \te \,\ote \,\,
{\rm Tr}\,\Biggl ( Q^\dagger e^{V} Q+ {\ov Q}^\dagger e^{-V} {\ov Q}\Biggr )+
\eq
\bq
+\int \te \,\Biggl \{-\frac{2\pi}{\alpha(\mu)}\,S+m_Q(\mu)\,{\rm Tr}\,Q{\ov Q}
\Biggr \}+\rm {h.c.}\,,\quad S=W_{\alpha}^2/32\pi^2\,, \nonumber
\eq
where $\alpha(\mu)$ is the running gauge coupling (with its scale parameter $\la$,
{\it independent of quark masses}), $m_Q(\mu)$ is the running current quark mass, $W_
{\alpha}$ is the gluon field strength, and traces are over color and flavor indices. This
theory has the exact $SU(N_c)$ gauge and, in the chiral limit $m_Q\ra 0$, global symmetries:
\bq
SU(N_F)_L\times SU(N_F)_R\times U(1)_B\times U(1)_R\,.\nonumber
\eq
Under these symmetries, the quarks $Q$ and $\ov Q$ transform as:
\bq
Q\,\,:\quad (N_c)_{\rm col}\times ( N_{F})_L^{\rm fl}\times (0)_R^{\rm fl}\times (1)_B\times
(N_F-N_c/N_F)_R\,,\nonumber
\eq
\bq
{\ov Q}\,\,:\quad ({\ov N}_c)_{\rm col}\times (0)_L^{\rm fl}\times ( {\ov N}_{F})_R^{\rm fl}
\times (-1)_B \times (N_F-N_c/N_F)_R\,.\nonumber
\eq
\vspace{0.2 cm}

The explicit dependence of the gluino condensate $\langle S \rangle $ on the current quark
masses and $\la$ can be found as follows.\\

a) One can start with $N_F < 3N_c$ and the heavy quarks, $m_Q^{\rm pole}\equiv m_Q(\mu=m_Q^
{\rm pole})\gg \la$\,, so that the theory is UV-free and in the weak coupling regime at
sufficiently large $\mu$.

b) Then, to integrate out all quarks directly in the perturbation theory at scales
$\mu < \mu_H=m_Q^{\rm pole}$,
resulting in the pure Yang-Mills theory with the scale factor $\Lambda_{YM}$.
The value of $\Lambda_{\rm YM}$ can be found from the matching of couplings $\alpha_{+}(\mu)
$ and $\alpha_{-}(\mu)$ of the upper and lower theories at $\mu=\mu_H\,:  \alpha_{+}
(\mu_H)=\alpha_{-}(\mu_H)$. The upper theory is always the original one with $N_c$ colors
and $N_F$ flavors, and the value of $\alpha_{+}(\mu_H)$ can be obtained starting with high
$\mu\gg \mu_H$ and evolving down to $\mu=\mu_H$ through the standard perturbative RG-flow for
theory with $N_c$ colors and $N_F$ flavors of massless quarks.
\footnote{\,
In (2.2),(2.3) and everywhere below in the text the perturbative NSVZ  \cite{NSVZ}
$\beta$\,-\,function is used, corresponding to the Pauli-Villars scheme.
}
But instead, the same value $\alpha_{+}(\mu)$ can be obtained starting with $\mu\sim\la$
and going up to $\mu=\mu_H$ with the same RG-flow for {\it massless} quarks. I.e.\,(${\rm g}
^2(\mu)=4\pi\alpha(\mu),\,\bo=3N_c-N_F$)\,:
\bq
\frac{2\pi}{\alpha_{+}(\mu_H)}=\bo\ln\,\Bigl (\frac{\mu_H}{\la}\Bigr )-N_F\ln \Bigl (\frac
{1}{{\it z}_Q(\mu_H,\,\la)}\Bigr )+ N_c\ln \Bigl (\frac{1}{{\rm g}^2(\mu_H)}\Bigr )+C_{+} \,,
\eq
where $z_Q=z_Q(\mu_H,\,\la)\ll 1$  is the standard perturbative renormalization factor
(logarithmic in this case) of massless quarks in theory with $N_c$ colors and $N_F$ flavors.

As for the lower theory, in all examples considered in this section it is the Yang-Mills
one with $N_c^{\,\prime}$ colors and no quarks. Its coupling can be written in a similar way as:
\bq
\frac{2\pi}{\alpha_{-}(\mu_H)}=3N_c^{\,\prime} \ln\,\Bigl ( \frac{\mu_H}{\Lambda_
{\rm YM}}\Bigr )+N_c^{\,\prime}\ln\Bigl (\frac{1}{{\rm g}^2(\mu_H)}\Bigr )+C_{-}\,.
\eq
$C_{\pm}$ in (2.2),(2.3) are {\it constants independent of the quark mass values}.
Our purpose here and everywhere below is to trace explicitly the dependence on
the parameters like $\mu_H/\la$ which will be finally expressed through the universal
parameter $m_Q/\la\,,\,\,m_Q\equiv m_Q(\mu=\la)$, which can be large $m_Q/\la\gg 1$,
or small $m_Q/\la\ll 1$. So, from now on and everywhere below the constant terms like
$C_{\pm}$ will be omitted, as their effect is equivalent to a redefinition of $\la$ by a
constant factor.
\footnote{\,
Introducing the Wilsonian coupling $\alpha_W(\mu)$ whose $\beta$\,-\,function is that of
NSVZ for $\alpha(\mu)$ but without the denominator, $2\pi/\alpha_W(\mu)=2\pi/
\alpha(\mu)-N_c\ln (1/{\rm g}^2(\mu))$\, \cite{NSVZ}, one has: $C_{+}=2\pi/\alpha_W^{+}
(\mu=\la),\,C_{-}=2\pi/\alpha_W^{-}(\mu=\Lambda_{YM})$. In essence, the term $N_c\ln
\Bigl (1/{\rm g}^2(\mu_H)\Bigr )$ in (2.2) is the higher loop perturbative renormalization
factor of gluons, i.e. $N_c\ln \Bigl (z_{\rm g}(\mu_{H}\,,\la)\Bigr )= N_c\ln \Bigl
(\alpha_{+}(\mu_{H})/\alpha_{+}(\mu=\la)\Bigr )$\,, and similarly in (2.3).
}

In the case considered now\,: $N_c^{\prime}=N_c\,,\,\mu_H=m_Q^{\rm pole}\gg \la $\,, and one
obtains then from (2.2),(2.3)\,:
\bq
\Lambda_{YM}=(\la^{\bo}\det m_Q)^{1/3N_c},\quad m_Q\equiv z_Q^{-1}(m_Q^{\rm
pole}\,, \la)\, m_Q^{\rm pole}\gg m_Q^{\rm pole}\gg \la\,.
\eq

c) Lowering the scale $\mu$ down to $\mu < \Lambda_{YM}$ and integrating out all gauge
degrees of freedom, except for the one whole field $S$ itself, one can write the effective
Lagrangian in the Veneziano-Yankielowicz (VY) form\, \cite{VY}, from which one obtains the
gluino condensate:
\bq
\langle S \rangle=\Lambda_{YM}^3=(\la^{\bo}\det m_Q)^{1/N_c}\,, \quad m_Q=m_Q(\mu=\la)\,.
\eq

Now, the expression (2.5) can be continued in $m_Q$ from large $m_Q\gg \la$ to small values,
$m_Q\ll\la$. While $m_Q$ for $m_Q\gg \la$ is some formally defined parameter\, (see (2.4), the
physical quark mass is $m_Q^{\rm pole}\gg \la$ and it does not run any more at $\mu< m_Q^{\rm
pole}$), at $m_Q\ll\la$ it has a simple and direct meaning: $m_Q=m_Q(\mu=\la)$.

The expression (2.5) for $\langle S \rangle$ appeared in the literature many times before, but
to our knowledge, the exact definition of the parameter $m_Q$ entering (2.5), i.e. its
relation with $m_Q(\mu)$ entering (1) which defines the theory, has not been given. Clearly,
without this explicit relation the expression (2.5) has no much meaning, as the quark mass
parameter $m_Q(\mu)$ is running. For instance, if $m_Q$ is understood as $m_Q^{\rm pole}$ in
(2.5) for heavy quarks, the relation $\langle S \rangle=(\la^{\bo}\det m_Q^{\rm pole})^{1/N_c}$
will be erroneous. All this becomes especially important, in particular, at $3N_c/2 <N_F< 3N_c$
and $m_Q\ll \la$,
when $m_Q(\mu)$ runs in a power-like fashion: $m_Q(\mu_2)=(\mu_1/\mu_2)^{\bo/N_F}\,m_Q(\mu_1)$.
Everywhere below, except for the section 8, only the case $m_Q\ll \la$ will be considered.\\

d) From the Konishi anomaly equation \cite{Konishi}\,:
\bq
\langle \, \Bigl ({\ov Q}_{\ov j}\,Q^i\Bigr )_{\mu} \,\rangle =\Bigl ( m^{-1}_Q (\mu)\Bigr )^
i_{\ov j}\,\langle S \rangle
\eq
one obtains the explicit value of the chiral condensate:
\bq
\langle ({\ov Q}_{\ov j}\,Q^i)_{\mu=\la}\rangle \equiv \cm^2\,\delta^i_{\ov j}=\frac{\langle S
\rangle}{m_Q}\,\delta^i_{\ov j}\,\,,\quad \cm =\Bigl(\la^{\bo}\,m_Q^{\nd}\Bigr )^{1/2N_c},
\quad \,\nd=N_F-N_c\,, \nonumber
\eq
\bq
\langle S \rangle=\Lambda^{3}_{YM}=(\la^{\bo}\det m_Q)^{1/N_c}\,,\quad m_Q\ll\la\,.
\eq

Now, the expression (2.5) can be continued in $N_F$ from the region $N_F<3 N_c$ to
$N_F> 3N_c$  and, together with the Konishi anomaly relation (2.6),  these two become then
the basic universal relations for any values of quark masses and any $N_F$.\\

To check this universal form of (2.5), let us consider briefly (see section 8 for more detail)
the case $N_F > 3N_c$ and $m_Q\ll \la$. In this case $\bo=(3N_c-N_F)<0$, so that the theory
is IR-free in the interval $\mu_H<\mu<\la$, where $\mu_H$ is the highest physical mass ($\lym
\ll\mu_H=m_Q^{\rm pole}\ll \la$ in this example). I.e., its coupling which is $O(1)$ at $\mu=
\la$ becomes
logarithmically small at $\mu\ll \la$. Besides, the parameter $m_Q$ has now a direct physical
meaning as the value of the running quark mass at $\mu=\la$,\, $m_Q\equiv m_Q(\mu=\la)\ll \la$.
So, starting with $\mu=\la$ and going down perturbatively to $\mu_H=m_Q^{\rm pole}=m_Q(\mu=m_Q^
{\rm pole})=z_Q^{-1}(\la,\, m_Q^{\rm pole})\,m_Q \gg m_Q$ (\,$z_Q(\la,\, m_Q^{\rm pole})\ll 1$
is the perturbative logarithmic renormalization factor of massless quarks), one can integrate
then out all quarks as heavy ones.  Writing the matching condition for two couplings $\alpha_{
+}$ and $\alpha_{-}$, one obtains (2.2),(2.3) with the only replacement: $z_Q(m_Q^{\rm pole}
\gg \la,\,\la)\ra z_Q^{-1}(\la,\, m_Q^{\rm pole}\ll \la)$, and the same expression (2.5).\\

Another check can be performed for $N_F < N_c-1$ and small quark masses, $m_Q\ll \la$. In this case
all quarks are higgsed and the gauge symmetry $SU(N_c)$ is broken down to $SU(N_c^{\prime}=N_c-N_F)$
at the high scale $\mu_H=\mu_{\rm gl}\gg \la$\,: $\langle Q^{i}_{a}\rangle_{\mu=\mu_
{\rm gl}}=\delta^{i}_{a}\co,\, \langle {\ov Q}_{\ov j}^{a}\rangle_{\mu=\mu_{\rm gl}}=\delta_
{\ov j}^{a}\co,\,\,\co\gg \la$. $(2N_c N_F-N_F^2)$ gluons become massive, with the mass
scale $\mu^2_{\rm gl}={\rm g^2_{+}}\langle{\hat\Pi}\rangle={\rm g}^2_{+}\co^2\,,\,\,{\rm g}^2_{+}=4\pi
\alpha_{+}(\mu=\mu_{\rm gl},\la)\ll 1$. The same number of quark degrees
of freedom acquire the same masses and become the superpartners of massive gluons (in a sense, they
can be considered as the heavy "constituent quarks"), and there remain $N_F^2$ light
complex pion fields  ${\hat \pi}^i_{\ov j}\,\,:\,{\hat \Pi}^i_{\ov j}=({\ov Q}_{\ov j}Q^i)_
{\mu=\mu_{\rm gl}}=\co^2(\delta^i_{\ov j}+{\hat \pi}^i_{\ov j}/\co),
\,\,\langle {\hat \Pi}^i_{\ov j}\rangle=\delta^i_{\ov j}\co^2$.

All heavy particles can be integrated out at scales $\mu<\mu_{\rm gl}$\,. The numerical matching of
couplings at $\mu_H=\mu_{\rm gl}$\,: $\alpha_{+}(\mu=\mu_{\rm gl},\la)$ in (2.2), i.e. those of the
original theory, with $\mu^2_{\rm gl}={\rm g}^2_{+}\langle{\hat\Pi}\rangle={\rm g}^2_{+}\co^2,\,\,
{\hat\Pi}=({\ov Q}Q)_{\mu=\mu_{\rm gl}}$\,, and $\alpha_{-}(\mu=\mu_{\rm gl},\Lambda_L)$
in (2.3) of the lower energy pure Yang-Mills
theory can be performed similarly to the previous examples with heavy quarks. But in this case we
consider it will be more useful to write the explicit form of the $\hat\Pi$ -dependence of the lower
energy coupling $\alpha_{-}(\mu<\mu_{\rm gl},\Lambda_L)$ multiplying the field strength squared
of massless gluons, to see how the multi-loop $\beta$ - function reconciles with the holomorphic
dependence of $\Lambda_L$ on the chiral superfields $\hat \Pi$\,. This looks now as :
\bq
\frac{2\pi}{\alpha_{-}\Bigl (\mu<\mu_{\rm gl},\Lambda_L\Bigr )}=
\Biggl \{3(N_c-N_F)\ln\,\Bigl (\frac{\mu}{\la}\Bigr )+(N_c-N_F)\ln \Biggl (\frac{1}{{\rm g}^2_
{-}(\mu,\langle\Lambda_L\rangle)}\Biggr )\Biggr \}+ \nonumber
\eq
\bq
+\Biggl \{\frac{3}{2}\ln \Biggl (\frac{{\rm g}^{2N_F}_{+}(\mu=\mu_{\rm gl},\la)\det {\hat \Pi}}{\la^
{2N_F}}\Biggr )+N_F\ln \Biggl (\frac{1}{{\rm g}^2_{+}(\mu=\mu_{\rm gl},\la)}\Biggr )\Biggr\}-
\nonumber
\eq
\bq
-\Biggl\{\frac{1}{2}\ln \Biggl (\frac{{\rm g}^{2N_F}_{+}(\mu=\mu_{\rm gl},\la)\det {\hat \Pi}}{\la^
{2N_F}}\Biggr )+N_F\ln \Biggl (\frac{1}{z_Q(\mu=\mu_{\rm gl},\,\la)}\Biggr )\Biggr\}\,,
\eq
where three terms in curly brackets in (2.8) are the contributions of, respectively, massless
gluons, massive gluons, and higgsed quarks.

It is worth noting that the dependence of the coupling $2\pi/\alpha_{-}$ on the quantum pion
superfields ${\hat\pi}^{i}_{\ov j}/\co$ entering ${\hat \Pi}^i_{\ov j}$ originates {\it only} from the
${\hat \pi}/\co$ -dependence of heavy particle masses entering the "normal" one-loop contributions to
the gluon vacuum polarization, while the "anomalous" higher loop contributions \cite{NSVZ} originating
from the quark and gluon renormalization factors $z_Q$ and $z^{\pm}_{\rm g}\sim {\rm g^2_{\pm}}$ (see
the footnote 2) do not contain the quantum pion fields $\hat\pi/\co$ and enter (2.8) {\it as pure
neutral c-numbers}. This is clear from the R-charge conservation (see the footnote 3) or, equivalently,
from {\it the holomorphic dependence of F-terms on chiral quantum superfields} (the chiral superfields
here are $\p (\mu_1)=z_Q(\mu_1,\,\mu_2)\,\p (\mu_2))$.

So, the coupling  $\alpha_{-}(\mu,\,\Lambda_L )$ of the lower energy
pure Yang-Mills theory at $\mu<\mu_{\rm gl}$ and its scale factor $\Lambda_L$ look as:
\bq
\hspace{-5mm}\frac{2\pi}{\alpha_{-}^{W}\Bigl (\mu<\mu_{\rm gl},\Lambda_L\Bigr )}=\frac{2\pi}{\alpha_
{-}\Bigl(\mu<\mu_{\rm gl},\Lambda_L\Bigr )}-(N_c-N_F)\ln\frac{1}{{\rm g}^2_{-}(\mu<\mu_{\rm gl},
\langle\Lambda_L\rangle)}=3(N_c-N_F) \ln\Bigl ( \frac{\mu}{\Lambda_L}\Bigr ),\nonumber
\eq
\bq
\Lambda_L^{3(N_c-N_F)}=\frac{\la^{\bo}}{z^{N_F}_Q(\mu_{\rm gl},\,\la)\det {\hat \Pi}}
\equiv \frac{\la^{\bo}}{\det {\Pi}}=\lym^{3(N_c-N_F)}\Biggl (\det\frac
{\langle \Pi\rangle}{\Pi}\Biggr )\,,
\eq
\bq
\Pi\equiv z_Q(\mu_{\rm gl},\,\la)\hat \Pi\,,\quad \langle \Pi\rangle=\cm^2\ll \co^2\,, \quad
\langle\Lambda_L\rangle=\lym=\Bigl (\la^{\bo}\det m_Q\Bigr )^{1/3N_c}\,,
\eq
and the Lagrangian at $\mu<\mu_{\rm gl}$ takes the form :
\footnote{\,
Because the gluon fields are not yet integrated completely, there are the gluon regulator
fields (implicit) whose contributions ensure the R-charge conservation in (2.11), see also
(2.12) below.

Besides, we neglected in (2.11) the additional dependence of the Kahler term on the quantum pion
fields $\hat \pi/\co$ (originating from the dependence on $\hat \pi/\co$ of the quark renormalization
factor $z_Q({\hat \Pi}^{\dagger},{\hat \Pi}$)\,), because at the weak coupling this will
influence the pion mass values through logarithmically small corrections only.
}
\bq
L=\int \te\ote  \Biggl \{2\,\rm {Tr}\sqrt {{\hat\Pi}^\dagger {\hat\Pi}}\Biggr \}+\int \te
\Biggl \{ -\frac{2\pi}{\alpha_{-}(\mu,\Lambda_L)}{\hat S}+ {\hat m_Q}\rm {Tr} {\hat\Pi} \Biggr \}\,,
\eq
where $\hat S={\hat W}^2_{\alpha}/32\pi^2$\,,\, and ${\hat W}_{\alpha}$ are the gauge field
strengths of $(N_c-N_F)^2-1$ remaining massless gluon fields.

Lowering the scale $\mu$ down to $\mu < \Lambda_{YM}$ and integrating out all gauge degrees
of freedom, except for the one whole field $\hat S$ itself (this leaves behind a large number
of gluonia with masses $M_{\rm gl}\sim \Lambda_{YM}$), one obtains the VY - form :
\bq
L=\int \te \,\ote \, \Biggl \{ 2\,\rm {Tr}\,\sqrt {{\hat\Pi}^\dagger {\hat\Pi}}
\,\,+({\rm D \,\,terms\,\, of\,\, the\,\, field  \,\,\hat S  \,})\Biggr \}+\nonumber
\eq
\bq
+\int \te \Biggl \{ -(N_c-N_F)\,{\hat S} \Biggl ( \ln \Bigl (\frac{\hat S}{\Lambda_{L}^3}
\Bigr )-1\Biggr )+{\hat m_Q}\rm {Tr}\, {\hat\Pi} \Biggr \}\,,\,\,\, \mu \lesssim \Lambda_{YM}\,.
\eq
It is worth noting that it is the first place where the non-perturbative effects were incorporated to
obtain the VY\,-\,form of the superpotential (the non-perturbative effects introduce the infrared
cutoff $\sim \lym$\,, so that the explicit dependence on $\mu$ disappears at $\mu<\lym$), while all
previous calculations with this example were purely perturbative.
One obtains from (2.12) the gluino vacuum condensate: $\langle \hat S\rangle=\langle
\Lambda_L^3\rangle=\Lambda_{YM}^3=\langle S\rangle=(\la^{\bo}\det m_Q)^{1/N_c}$\,.

Finally, integrating out the last gluonium field $\hat S$ (with its mass scale $\sim\Lambda_{YM}$\,)
at lower energies, one obtains the Lagrangian of pions:
\bq
L=\Biggl [2\,\rm {Tr}\,\sqrt {{\hat\Pi}^\dagger {\hat\Pi}}
\Biggr ]_D+\Biggl [ (N_c-N_F)\Biggl (\frac{\la^{\bo}}{{\it z}_Q(\mu_{\rm gl},\,\la)\det\,\hat\Pi}
\Biggr )^{1/(N_c-N_F)}+ {\hat m}_Q\rm {Tr}\, {\hat\Pi} \Biggr ]_F=
\eq
\bq
=\Biggl [\frac{2}{{\it z}_Q(\mu_{\rm gl},\,\la)}\,\rm {Tr}\,\sqrt {{\Pi}^\dagger {\Pi}}
\Biggr ]_D+\Biggl [ (N_c-N_F)\Biggl (\frac{\la^{\bo}}{\det\,{\Pi}}
\Biggr )^{1/(N_c-N_F)}+ {m}_Q\rm {Tr}\, {\Pi} \Biggr ]_F\,,\quad \mu\ll\Lambda_{YM}\,.\nonumber
\eq

The superpotential of the form $(N_c-N_F)({\la^{\bo}}/\det\,{\Pi})^{1/\nd}$ appeared many times in the
literature because, up to an absolute normalization of the field $\Pi$ (which is not RG-invariant by 
itself), this is the only possible form of the superpotential, if one is able to show that the lowest 
energy Lagrangian depends on $N_F^2$ pion superfields only. But it seems, the absolute
normalization of all terms entering (2.13) has never been carefully specified (clearly, the absolute
normalization makes sense only when both the superpotential and the Kahler terms are absolutely
normalized simultaneously). The Lagrangian (2.13) describes weakly interacting pions with small masses
$M_{\pi}=2 {\hat m}_Q=2 z_Q(\mu_{\rm gl},\,\la) m_Q\ll m_Q\ll \Lambda_{YM}\ll \la$\,.\\

On the whole, the mass spectrum contains in this case: $(2N_c N_F-N_F^2)$ massive gluons and
"constituent quarks" with the mass scale $\mu_{\rm gl}={\rm g}_H\co\gg \la$, a large number of
gluonia with the mass scale $\sim \Lambda_{YM}\ll \la$\,, and $N_F^2$ pions with small masses
$M_{\pi}= 2{\hat m}_Q=2m_Q(\mu=\mu_{\rm gl})\ll \Lambda_{YM}$.

The form (2.13) can be continued in $N_F$ to the point $N_F=N_c-1$ and it predicts then the form
of the pion Lagrangian for this case. Now, the whole gauge group is higgsed at the high scale
$\mu_H=\mu_{\rm gl}\gg \la$, and the direct way to obtain (2.13) is not through the
VY\,-\,procedure, but through the calculation of the one-instanton contribution
\cite{ADS}\cite{SV-r}. The changes in the mass spectrum are evident and, most important,\,-\,
there is no confinement and there are no particles
with masses $\sim\Lambda_{YM}$ in the spectrum in this case. \\

\section {Direct theory\,.\,\,Conformal window\\
{\*\hspace {2.5 cm}\boldmath $3N_c/2 <N_F <3N_c$}}

\hspace {6 mm} The superconformal behavior means the absence of the scale $\la$ in the physical mass
spectrum. In other words, there are no particles with masses $\sim \la$, all quarks and gluons remain
effectively massless at $\mu_H\ll\mu\ll \la$, where $\mu_H$ is the highest physical mass scale.
So, "nothing especially interesting" happens when decreasing the scale $\mu$ from $\mu\gg\la$
down to $\mu_H\ll\mu<\la$. Only the character of running of the coupling $\alpha(\mu)$ and
the quark renormalization factor $z_Q(\mu)$ change. The slow logarithmic evolution in the
weak coupling region $\mu\gg \la$ is replaced by freezing of $\alpha(\mu)$ at $\mu <
\la$\,:\,\,$\alpha(\mu)\ra \alpha^{*}$\,,\, while $z_Q(\mu)$ acquires the power behavior\,:\,\,
 $z_Q(\la,\mu)=(\mu/\la)^{\bo/N_F}< 1$. As a result, the Green functions of chiral superfields
also behave in a power-like fashion, with dynamical dimensions determined by their R-charges\,:
$\rm D=3|R|/2$. This conformal regime continues until $\mu$ reaches at $\mu\ll \la$ the highest
physical mass scale $\mu_H\ll \la$, and then the conformal behavior breaks down.\\

There are three characteristic scales at $\mu=\la$ in the direct theory\,: the current quark
mass $m_Q$, the scale $\cm$ of its chiral vacuum condensate, and the scale $\Lambda_{YM}$ of
the gluino condensate. It is seen from (2.5-2.7) that in the whole region $N_c <N_F < 3N_c$
there is an hierarchy:
\bq
m_Q\ll \Lambda_{YM}\ll \cm\,\quad {\rm at} \quad N_c <N_F < 3N_c\,.
\eq

By itself, this hierarchy has no direct physical consequences, until it is realized that some
{\it physical masses} stay behind the above quantities. As will be shown below, within the
dynamical scenario considered, the above inequalities reflect a real hierarchy
of physical masses\,: $m_Q$ is the mass of lightest pions, $\lym$ is the mass scale of gluonia
and $\cm$ is the dynamical constituent mass of quarks.
\\

The main idea of the dynamical scenario for SQCD, with $N_c <N_F < 3N_c$ and small equal quark
masses, considered in this paper is that this theory is in the collective coherent
"diquark-condensate" (DC) phase. This means that quarks
don't condense alone, $\langle Q^i\rangle=\langle {\ov Q}_{\ov j}\rangle=0$\, (because there
are too many flavors at $N_f>N_c$). In other words, theory is not higgsed by quarks, all
gluons remain massless at scales $\mu\gg \lym$, and color is confined. But quarks condense in
colorless chiral pairs $({\ov Q}_{\ov j}Q^i)$ and these pairs form the {\it coherent} condensate
( like the quark-antiquark pairs in the Nambu-Jona-Lasinio model and, more importantly, like QCD).
And as a result of this coherent condensation, quarks acquire large (in comparison with their pole
mass, $m_{Q}^{\rm pole}=m_Q(\mu=m_{Q}^{\rm pole})$\,) dynamical constituent mass $\mu^{2}_C=
\langle \Pi_{2}\rangle=\langle\,{\Bigl ({\ov Q} Q\Bigr )}_{\mu=\mu_2}\,\rangle$\,,\,\,(\,
$\mu_2=\mu_C/(\rm several)\,,\, \mu_C=\cm$\,, see below). This constituent quark mass $\mu_C=\cm$
is the highest physical mass $\mu_H$ and it stops the massless perturbative RG-evolution at scales
$\mu<\mu_C$. Simultaneously, the light composite pions $\pi^i_{\ov j}$ are formed, with masses
$M_{\pi}\sim m_2=m_Q(\mu=\mu_2)$\,,\, ($m_2=m_Q$\,, see below).
\footnote{\,
This is unlike (our) QCD, where the value of the constituent quark mass $\mu_C$ is also determined by the
coherent chiral quark condensate, $\mu_C^3=\langle \psi{\ov \psi}\rangle$, but it is here $\mu_C\sim \la$\,,
while $m_{\pi}\sim (m_Q \,\mu_C)^{1/2}$\,. The difference in parametrical dependence of $m_{\pi}$ on
the current quark mass $m_Q$ between SQCD and QCD
is because the spin $1/2$ quarks are condensed in QCD, while these are spin zero quarks in SQCD.

Besides, unlike the genuine spontaneous breaking of the chiral flavor symmetry in QCD with $\mu_C
=\langle{\psi\ov \psi}\rangle ^{1/3}\sim \la\neq 0$ at $m_Q\ra 0$, in SQCD $\mu_C=\langle {\ov Q}Q_{
\mu=\la}\rangle^{1/2}=\cm\ra 0$ at $m_Q\ra 0$, see (2.7). Nevertheless, because the ratio $\cm/
m_Q\gg 1$ is parametrically large at $m_Q\ll \la$, all qualitative features remain the same, so
that this can be considered as the "quasi-spontaneous breaking" of the chiral flavor symmetry.
}

All this occurs in the "threshold region" $\mu_2=\mu_C/({\rm several})<\mu <\mu_1=
({\rm several})\,\mu_C$ around the scale $\mu_C$ of the constituent quark mass.
In other words, the non-perturbative effects operate in this threshold region, so that they
"turn on" at $\mu=\mu_1$ and "saturate" at $\mu=\mu_2$.

If this idea is accepted, the proposed effective Lagrangian at the scale $\mu_2$ has the form:
\bq
L=\int \te \,\ote \, \Biggl \{ \rm {Tr}\,\sqrt {\Pi^\dagger_2 \Pi_2}+Z_2{\rm Tr}\Biggl ( Q
^\dagger_2 e^{V} Q_2+ {\ov Q_2}^\dagger e^{-V} {\ov Q_2}\Biggr ) +\cdots\Biggr \}+\nonumber
\eq
\bq
+\int \te \,\Bigl ( {\rm W_g}+W_Q \Bigr ) +\rm {h.c.}\,,\quad
 W_g=-\frac{2\pi}{\alpha(\mu_2)}\,S\,,\quad S=W_{\alpha}^2/32\pi^2\,,
\eq
\bq
W_Q=\Biggl ( \frac{\rm {\det}\,\Pi_2}{\la^{\bo}}\Biggr )^{1/\ov N_c}
\rm {Tr} \Bigl (\ov Q_2 \,\Pi_2^{-1}\, Q_2\Bigr )\,
-N_F \Biggl ( \frac{\rm {\det}\,\Pi_2}{\la^{\bo}}\Biggr )^{1/\ov N_c}+
m_2\,\rm {Tr}\, \Pi_2\,,\,\nonumber
\eq
\bq
Z_2=\frac{\Lambda_o}{\mu_C}=\Biggl (\frac{\mu_C}{\la}\Biggr )^{\bo/\nd}=\frac{m_Q}{\cm}\,,
\quad\Lambda_o=\frac{1}{\langle \Pi_2 \rangle}\Biggl ( \frac{\rm {\det}\,\langle\Pi_2
\rangle}{\la^{\bo}}\Biggr )^{1/\ov N_c}\,,\quad \nd=N_F-N_c\,.\nonumber
\eq

Here, the field $(\Pi_2)^i_{\ov j}=\Bigl ({\ov Q_{\ov j}}\,Q^i\,\Bigr )^{\rm (light)}_{\mu=\mu_
2}$ represents the dynamically generated "one-particle light part" of the composite field\,:
$(\Pi_2)^i_{\ov j}=\mu^2_C \Bigl (\delta^i_{\ov j}+\pi_{\ov j}^i/\mu_C\Bigr )$, it contains
the c-number vacuum part $\mu_C^2\,\delta^i_{\ov j}=\langle \,(\Pi_2)^i_{\ov j}\,\rangle=
\langle \,{\ov Q}_{2,\,\ov j}\,Q_2^{i}\,\rangle\equiv \langle \,{\ov Q}_{\ov j}\,Q^{i}\,
\rangle_{\mu=\mu_2}$\,, and the quantum fields $\pi^i_{\ov j}/\mu_C$ of light pions .
The canonically normalized quark
fields $C_2=Z_2^{1/2}Q_2$ and ${\ov C}_2=Z_2^{1/2}\,{\ov Q}_2$ have no c-number vacuum parts,
$\langle C\rangle=\langle \ov C\rangle=0$, and are the quantum fields of heavy
constituent quarks with "the field masses" $(\mu_C)_i^{\ov j}$ and c-number masses $\mu_C$:
\bq
(\mu_C)_i^{\ov j}=\frac{1}{Z_2}\Biggl ( \frac{\det\,\Pi_2}{\la^{\rm {b_o}}}\Biggr )^{1/{\ov
N_c}}\,\Bigl (\Pi_2^{-1}\Bigr )_i^{\ov j}\,,\quad \langle (\mu_C)_i^{\ov j}\rangle =
\delta_i^{\ov j}\,\mu_C\,\,.
\eq

The nonzero vacuum condensate $\langle {\ov C_{2,\,\ov j}}C_2^i\rangle =Z_2
\langle \, {\ov Q_{2,\ov j}}\,Q_2^i\,\rangle=Z_2\,\mu_C^2\,\delta^i_{\ov j}=\bigl (\langle
S\rangle /\mu_C \bigr )\,\delta^i_{\ov j}$ of these heavy constituent quarks is a pure
quantum effect from the one-loop triangle diagram with the constituent quark fields
$C_2$ and ${\ov C}_2$ contracted into their massive propagators with the masses $\mu_C$
and emitting two external gluino lines, this contribution realizes the Konishi anomaly.

Besides, by definition, all effects of evolution through the threshold region are already
taken into account in (3.2), so that the quark terms in the Lagrangian are needed practically
for calculations with the valence heavy quarks only. And finally, dots in (3.2) indicate
other possible D-terms which are supposed to play no significant role in what follows.

To a large extent, the form of the Lagrangian in (3.2) is unique, once the main assumption
about formation at the scale $\mu\sim \mu_C$ of massive constituent quarks with masses $\mu_C
^2=\langle {\ov Q}_2Q_2 \rangle=\langle \Pi_2\rangle$ and light pions with masses $m_2$ (and
with all gluons remaining massless) is adopted.
The only important non-trivial point, may be, is the non-zero value of the coefficient
$(-N_F)$ in front of the second term in the superpotential $W_Q$. This was determined
from the requirement that, until quark and/or gauge degrees of freedom are not
integrated out, the vacuum value of the superpotential is not changed yet, in
comparison with its original value at higher scales $\mu \gg \mu_C$: $\langle W_Q\rangle=\sum_
{\rm {flav}} m_Q(\mu)\langle (Q \ov Q)_{\mu}\rangle=N_F\langle S\rangle $ (contributions
of all three terms in $W_Q$ in (3.2) to $\langle W_Q\rangle$ are equal $N_F\langle S\rangle $
each, but the vacuum averages of the first and second terms in (3.2) cancel each other).

The absolute value of $\langle \Pi_2 \rangle=\langle {\ov Q}_2Q_2 \rangle$ can be determined
from the Konishi anomaly\,:
\bq
\frac{1}{\langle \Pi_2 \rangle}\Biggl ( \frac{\rm {\det}\,\langle\Pi_2\rangle}{\la^{\bo}}
\Biggr )^{1/\ov N_c}\langle {\ov Q}_2Q_2\rangle =\langle S\rangle=\Biggl ( \frac{\rm {\det}\,
\langle\Pi_2\rangle}{\la^{\bo}}\Biggr )^{1/\ov N_c}\,.
\eq
Together with $ m_2\langle \Pi_2\rangle=\langle S\rangle$\,, it follows from (3.4)
(see (2.5-2.7)) that\,:
$m_2=m_Q\equiv m_Q(\mu=\la)$ and $\mu_C^2=\langle
\Pi_2\rangle=\langle {\ov Q}_2Q_2\rangle=\langle {\ov Q}Q\rangle_{\mu=\la}\equiv \cm^2$.
\footnote{\,
It is worth noting that the concrete form of the Kahler term $K_{\pi}$ of quantum pion fields $
\pi^i_{\ov j}$ in (3.2) should not be taken literally. Its only purpose is to show a typical
scale of this Kahler term. For instance, one can replace it with the contribution $\sim {\rm Tr}\,
\Bigl({\mu_C}^{\dagger}\mu_C\Bigr)$ from the loop of constituent quarks, where the field
$(\mu_C)^{\ov j}_i$ is given in (3.3). Finally, to determine the values of pion masses up to
non-parametrical factors $\sim 1$\,, it is only important that both these forms of the pion Kahler
term have the same scale $\langle K_{\pi}\rangle\sim \cm^2$\,. For similar reasons, we neglect
possible additional dependence of $Z_2$ - factors entering the Kahler term of the constituent
quark in (3.2) on the quantum pion fields $\pi/\cm$.
}
\vspace{1mm}

It is also useful to consider the evolution through the threshold region in more detail. At
the scale $\mu=\mu_1$, there is no real distinction yet between the original light quarks
$Q_1=Q(\mu=\mu_1)$ and ${\ov Q}_1={\ov Q}(\mu=\mu_1)$ with the current masses $m_1=m_Q(\mu=
\mu_1)$ and the (heavy at scales $\mu<\mu_2$) constituent quarks $C_1=C(\mu=\mu_1),\,{\ov C}_1
={\ov C}(\mu=\mu_1)$, because the large constituent quark mass $\mu_C$
"turns on and saturates" only after the evolution through the threshold
region $\mu_2<\mu<\mu_1$. Similarly, there is no real distinction between the light
composite field $(Q{\ov Q})(\mu=\mu_1)$ with its mass scale $\sim m_1$ and the pion
field $\Pi_1=\Pi(\mu=\mu_1)$ (this is the pion $\Pi_2=\Pi(\mu=\mu_2)$ evolved back to
$\mu=\mu_1)$, with its mass $m_2$ at $\mu=\mu_2$ evolving back to the current quark mass
$m_1$ at $\mu=\mu_1$. In essence, all these are the obvious matching conditions. They can be
also used as an independent check that the form of $W_Q$ in (3.2) is self-consistent. After
evolving back from $\mu=\mu_2$ to $\mu=\mu_1$, the difference between the composite field ${\ov Q}Q$
of heavy constituent quarks and the field $\Pi$ of the light pion disappears due to disappearance
of the mass gap $\sim \mu_C$, so that two first terms in $W_Q$
cancel each other, while  the last term evolves back into the original quark mass term.

But then, at $\mu<\mu_1$, the colorless light composite pions and colored heavy constituent
quarks evolve differently
through the threshold region $\mu_2\leq \mu \leq \mu_1 $, and their Kahler terms acquire
different renormalization factors. The renormalization factor $Z_{\pi}$ of pions is: from
$\Pi_1\sim (Q_1{\ov Q}_1)$ with the mass $m_1$ at $\mu=\mu_1$ to $\Pi_2=Z_{\pi}\,\Pi_1,$ with
the mass $m_2$ at $\mu=\mu_2$, i.e.: $Z_{\pi}=m_1/m_2$. Similarly, the overall
renormalization factor of quarks is: from $({C_1}^{\dagger}C_1)\sim ({Q_1}^{\dagger}Q_1)$
with the mass $m_1$ at $\mu=\mu_1$ to $({C_2}^{\dagger}C_2)=Z_Q\,({C_1}^{\dagger}C_1)$,
with the mass $\mu_C$ at $\mu=\mu_2$\,, i.e.\,: $Z_Q=m_1/\mu_C$.

Independently of (3.4), the absolute values of $m_2$ (the parameter
$m_2$ will enter explicitly the lowest energy Lagrangian and will determine the
observable pole masses of pions, $M_{\pi}\sim m_2$) and $\langle \Pi_2\rangle=\mu_C^2$ can be
obtained from the following reasoning. Let us rewrite, say, the second term in the quark
superpotential in (3.2) in terms of the quark fields $(Q_1{\ov Q}_1)$ normalized at $\mu=\mu
_1$ and then, once more, in terms of $(Q_{\mu}{\ov Q}_{\mu})$ normalized at running
$\mu> \mu_1$:
\footnote{\,
It is worth noting that this is only a change of notations, not a real evolution to another scale.
}
\bq
\Biggl ( \frac{\det\,\Pi_2}{\la^{\bo}}\Biggr )^{1/\ov N_c}=
Z_{\pi}^{\, N_F/\nd}\Biggl ( \frac{\det\,(Q_1{\ov Q_1})}{\la^{\bo}}\Biggr )^{1/\ov N_c}=
\eq
\bq
\Biggl (Z_{\pi}z_Q(\mu,\mu_1) \Biggr )^{\, N_F/\nd}\Biggl ( \frac{\det\,(Q_{\mu}
{\ov Q}_{\mu})}{\la^{\bo}}\Biggr )^{1/\ov N_c}=
\Biggl (Z_{\pi}z_Q(\la,\mu_1) \Biggr )^{\, N_F/\nd}\Biggl ( \frac{\det\,(Q_{\la}
{\ov Q}_{\la})}{\la^{\bo}}\Biggr )_{\,.}^{1/\ov N_c}\nonumber
\eq

Clearly, at running $\mu_1\leq \mu\le \la$, the coefficient in front of the field
$(Q_{\mu}{\ov Q}_{\mu})$ depends explicitly on the running scale $\mu$ through the quark
perturbative renormalization factor $z_Q(\mu,\,\mu_1)$, while $Z_{\pi}$ is independent of
$\mu$. So, to find the value of $Z_{\pi}$, we have to fix the normalization at some definite
value of $\mu$. The only distinguished point is $\mu=\la$ in a sense that this term in the
superpotential, being expressed though the fields $(Q_{\mu=\la}{\ov Q}_{\mu=\la})$
normalized at $\la$, should have the coefficient which depends on $\la$ only.
From this, it follows:
\bq
Z_{\pi}=\frac{ m_1=m_Q(\mu=\mu_1)}{m_2}=z^{-1}_Q(\la,\,\mu_1)\equiv z^{-1}_Q\gg 1\,,\quad
\mu_1\sim \mu_C \ll \la\,,\nonumber
\eq
\bq
\mu^2_C=\langle \Pi_2\rangle=\langle {\ov Q}_2 Q_2\rangle=\langle \,{\ov Q}Q \,
\rangle_{\mu=\la}\equiv \cm^2\,,\quad m_2=m_Q(\mu=\la)\equiv m_Q\,,
\eq
\bq
Z_2=\frac{\Lambda_o}{\mu_C}=\Biggl (\frac{\cm}{\la}\Biggr )^{\bo/\nd}=\frac{m_Q}{\cm}\,,
\quad Z_Q=\frac{m_1}{\mu_C}=\frac{m_2}{\mu_C}\,\frac{m_1}{m_2}=\frac{m_Q}{\cm}\frac{m_1}
{m_2}=Z_2\,Z_{\pi}=Z_2 z^{-1}_Q\,,\nonumber
\eq
where $z_Q(\la,\,\mu=\mu_1)\ll 1$ is the standard perturbative renormalization factor
of the massless quark describing its evolution from $\mu=\la$ down to $\mu=\mu_1 $
( in the conformal window it is
known explicitly: $z_Q=z_Q(\la,\,\mu_1)=(\mu_1/\la)^{\bo/N_F}\ll 1 $ \,).

On the whole, the evolution of the current quark mass in the interval $\mu_2 \leq \mu\leq \la$
looks as follows. At $\mu=\la$ the current quark mass is $m_Q\equiv m_Q(\mu=\la)$. At
smaller $\mu$ it runs with the perturbative $z_Q^{(\mu)}=z_Q(\la,\,\mu)$ -factor, $m_Q(\mu)=
m_{\la}/z_Q^{(\mu)}\gg m_Q$, so that $m_1\equiv m_Q(\mu=\mu_1)=z^{-1}_Q m_Q$. In the threshold
region $\mu_2 < \mu <\mu_1$ it runs so that (~at $\mu<\mu_1$ the current quark mass can be
understood more properly as the pion mass)\,:\,
$m_1\equiv m_Q(\mu=\mu_1)\ra m_2\equiv m_Q(\mu=\mu_2),\, m_2=Z^{-1}_{\pi}\,m_1$. And at
$\mu \ll \mu_2$ the current quark mass $m_2$ does not run any more. Using that $Z_{\pi}=z_Q^
{-1}$ from (3.6), it is seen that, evolving through the threshold region from $\mu=\mu_1$ down
to  $\mu = \mu_2$, the current quark mass returns back to its value at $\mu=\la \,: m_2=Z_{\pi}
^{-1}\,m_1=Z_{\pi}^{-1}\,(z_Q^{-1}\,m_Q)=m_Q$. As for the constituent quark mass $\mu_C$\,, it
originates in the threshold region $\mu\sim \mu_C$ due to an existence of the {\it coherent}
quark condensate, $\mu^2_C=\langle {\ov Q}_2 Q_2\rangle=\cm^2$\,, and it stops the further
RG-evolution of the constituent quark and pion fields at $\mu<\mu_C=\cm$. The self-consistency
of this scenario requires that $\mu_C=\cm$ be larger than $m_Q^{\rm pole}$, because otherwise
the massless conformal regime will stop before at $\mu=m_Q^{\rm pole}$\,, i.e. quarks will be in
the HQ (heavy quark) phase and the coherent quark condensate can't be formed in this case.
In the case considered, with $3N_c/2<N_F<3N_c$\,,
\bq
\frac{m_Q^{\rm pole}}{\la}\equiv \frac{m_Q(\mu=m_Q^{\rm pole})}{\la}=\frac{m_Q}{\la}\Biggl
(\frac{\la}{m_Q^{\rm pole}} \Biggr )^{\bo/N_F}=\Biggl ( \frac{m_Q}{\la}\Biggr )^{N_F/3N_c}=
\frac{\lym}{\la}\ll \frac{\mu_C}{\la}=\frac{\cm}{\la}\,,\nonumber
\eq
so that this is self-consistent.\\

Let us dwell now on the evolution of the Wilsonian coupling $\alpha_W(\mu)$
in the interval $\mu_2 < \mu <\la$. Let us recall first its standard perturbative evolution
in the interval $\mu_1< \mu <\la$:
\bq
\delta \Biggl (\frac{2\pi}{\alpha_W(\mu)}\Biggr )=
\Biggl \{ 3N_c \ln \frac{\mu}{\la}-N_F\ln \frac{\mu}{\la}
\Biggr \}+\Biggl \{ N_F\ln \frac{1}{z_Q(\mu)} \Biggr \}\,,
\eq
where the first two terms are the one-loop contributions of massless gluons and quarks,
while the last term describes higher-loop effects from massless quarks \cite{NSVZ}. In
the conformal window $3N_c/2 < N_F <3N_c$ the explicit form of the quark renormalization
factor $z_Q(\mu)$ is known at $\mu < \la$: $z_Q(\mu)\equiv z_Q(\la,\mu)=(\mu/\la)^{\rm{b_o}
/N_F}\ll 1$. Then, the above three parametrically large logarithmic terms in (3.7) cancel each
other. This describes the standard effect that the perturbative coupling freezes in the
conformal regime at $\alpha^*=O(1)$, i.e. it remains nearly the same as it was at $\mu= \la
$\,, as $\alpha(\mu=\la)$ is already close to $\alpha^*$\,, by definition of $\la$\,.

This perturbative form (3.7) can be used down to $\mu >  \mu_1$. Now, on account of additional
contributions from the threshold region $\mu_2 <\mu < \mu_1$, the coupling $\alpha(\mu,\Lambda_L)$
at $\mu<\mu_2$ looks as (the number $2\pi/N_c\alpha(\mu=\la)$ is considered as $O(1)$ and is
neglected in comparison with the large logarithm):
\bq
\frac{2\pi}{\alpha_{W}(\mu<\mu_2,\Lambda_L)}=\Biggl \{\frac{2\pi}{\alpha(\mu<\mu_2,\Lambda_L)}-
N_c\ln\Biggl (\frac {1}{{\rm g^2}(\mu,\langle\Lambda_L\rangle)}\Biggr )\Biggr \}= \nonumber
\eq
\bq
=\Biggl \{3 N_c \ln\frac{\mu}{\la} -\ln\Biggl (\frac{\det \, (\mu_C)_i^{\ov j}}{\la^{N_F}}
\Biggr )+N_F\,\Biggl ( \ln \frac {1}{z_Q}+ \ln \frac{1}{Z_Q}\Biggr )\,\Biggr \}\,.
\eq
Here:

a)\, the first term in the curly brackets in (3.8) is due to contributions of massless gluons\,; b)\,
in the
second term in the curly brackets the one-loop term from colored quarks stops now its evolution at
their constituent mass $(\mu_C)^i_{\ov j}\,\,,\,\,$ see (3.3), i.e. with surviving light pion fields
$\pi^i_{\ov j}$ still living at lower energies; besides, in addition to the previous term $\ln(1/z_Q),
\, z_Q\equiv z_Q(\la,\mu_1)$\,, which describes the standard smooth perturbative evolution from $\mu=
\la$ down to $\mu_1$\,, there appeared the last term $\ln(1/Z_Q)$ which is due to the additional
(non-standard) evolution of
the colored constituent quark in the threshold region $\mu_2\leq \mu\leq \mu_1$.

Numerically (i.e. neglecting the quantum pion fields $\pi^i_{\ov j}/\cm$ and replacing $\det\Pi_2$ by
its vacuum value $\cm^{2N_F}$), the first three terms in the r.h.s. of (3.8) still cancel each other.
So, the parametrically large value of $1/\alpha_W(\mu<\mu_2)$ (i.e. the weak coupling) originates
from the parametrically large $\ln(1/Z_Q)$ threshold contribution only. In other words, the
strong evolution of the coupling $\alpha(\mu)$ in the threshold region $\mu_2<\mu<\mu_1$
decreases it from the $O(1)$ value at $\mu=\mu_1$ to a logarithmically small value $\alpha
(\mu_2)\sim\alpha_W(\mu_2)\sim 1/\ln(\la/\cm)$ at $\mu=\mu_2$.

Substituting into (3.8) the value of $Z_Q$ from (3.6) and $\det\, (\mu_C)_i^{\ov j}$ from (3.3),
one can write finally the Yang-Mills coupling as  :
\bq
\frac{2\pi}{\alpha_{W}(\mu<\mu_2)}=\Biggl \{\frac{2\pi}{\alpha(\mu,\Lambda_L)}- N_c\ln \frac{1}
{g^2(\mu,\langle\Lambda_L\rangle)}\Biggr \}=3N_c \ln \frac{\mu}{\Lambda_L}\,\,\,\,,\nonumber
\eq
\bq
\Lambda_L=\Biggl (\frac{\det \Pi_2}{\la^{\bo}}\Biggr )^{1/{3\nd}}\,,\quad
\Lambda_{YM}\equiv \langle \Lambda_L \rangle =\Biggl (\frac{\cm^{2N_F}}{\la^{\bo}}
\Biggr )^{1/3\nd}\,.
\eq

Let us emphasize (this will be important for us in section 7) that {\it the explicit value of
the quark perturbative renormalization factor $\,\,\,{ z_Q=z_Q(\la,\mu_1\sim\cm)}\,\,$ is
not really needed to obtain $(3.9)$, because ${ z_Q}$ cancels exactly in
$(3.8)$, independently of its explicit form (and $ Z_2$ also)}.\\

Now, at lower scales $\mu < \mu_2$\,, if we are not interested in calculations with the
valence quarks, the fields of heavy constituent quarks can be integrated out.
\footnote{\,
Because quarks are confined, this leaves behind a large number of heavy quarkonia,
both mesons and baryons, with masses $M_{\rm meson}\sim \cm$ and $M_{\rm baryon}\sim
N_c\,\cm$, built from {\it non-relativistic (and weakly confined, the string tension is
$\sqrt \sigma\sim \lym\ll \cm$)} constituent quarks with masses $\mu_C=\cm$. Indeed,
the characteristic distance between the non-relativistic quarks in the bound state is the
Bohr radius: $R_B\sim 1/p_B$, where $p_B$ is the Bohr momentum $p_B\sim \alpha({\ov \mu}
\simeq p_B)\,\cm$. Supposing that $p_B \ll \cm$, this requires $\alpha({\ov \mu} \ll
\cm)\ll 1$. But indeed (see above), in this region $\Lambda_{YM}\ll\mu \ll
\cm$ the coupling is already
logarithmically small, $\alpha(\mu)\sim 1/\ln(\mu/\Lambda_{YM})\ll 1$. So, the
nonrelativistic regime is self-consistent ($\alpha(\mu)$ becomes $O(1)$ only at much
smaller distances $R_{\rm ch}\sim 1/\cm\,\ll R_B$, while confinement effects begin to be
important only at much larger distances $R_{\rm conf}\sim 1/\Lambda_{YM}\gg R_B$).
}

This will result in simply omitting in (3.2) all terms containing the quark fields (let us
recall that the quark loop contributions to the gauge coupling have been taken into account
already in (3.8)). Besides, the pion fields $\Pi_2$ (and masses $m_2$) do not evolve any more at
$\mu<\mu_2$\,, so that $\cm$ from ${\cm}^2=\langle \Pi_2 \rangle $  and $m_2$
become the low energy constant observables   at $\mu\ll \cm$ (the pion pole mass will be $\sim
m_2$ and $\cm=\langle S\rangle /m_2$\,, or $\langle S\rangle$ itself, are connected with
tensions of BPS domain walls between different vacua \cite{DS}). Therefore, the only remaining
evolution in the interval $\Lambda_{YM}\ll \mu \ll \cm$ is the standard (weak coupling)
perturbative logarithmic evolution of massless gluons, so that in this range of scales the
Lagrangian takes the form (from now on, to simplify the notations, we substitute: $\Pi_2
\equiv \Pi$,\, and $m_2=m_Q\equiv m_Q(\mu=\la)$\,, see also the footnote 3 about the R-charge) :
\bq
L=\int \te\ote \Biggl \{ \rm {Tr}\sqrt {\Pi^\dagger \Pi}\Biggr \}
+\int \te \Biggl \{ -\frac{2\pi}{\alpha(\mu,\Lambda_L)} S
-N_F \Biggl ( \frac{\rm {\det}\Pi}{\la^{\bo}}\Biggr )^{1/\ov N_c}+
m_Q\rm {Tr} \Pi \Biggr \}\,,
\eq
\bq
 \Lambda_{L}=\Biggl ( \frac{\det\,
\Pi}{\la^{\bo}}\Biggr )^{1/{3\ov N_c}},\quad \Lambda_{YM}\ll \mu \ll \cm\,.
\eq

Lowering the scale $\mu$ down to $\mu < \Lambda_{YM}$ and integrating out all gauge degrees
of freedom, except for the one whole field $S$ itself (this leaves behind a large number of
gluonia with masses $M_{\rm gl}\sim \Lambda_{YM}$), one obtains the VY\,-\,form :
\bq
L=\int \te \,\ote \, \Biggl \{ \rm {Tr}\,\sqrt {\Pi^\dagger \Pi}\Biggr \}\,\,+
({\rm D \,\,terms\,\, of\,\, the\,\, field  \,\,S  \,})+\nonumber
\eq
\bq
+\int \te \Biggl \{ -N_c\, S \Biggl ( \ln \frac{S}{\Lambda_{L}^3}-1\Biggr )
-N_F \Biggl ( \frac{\det\,\Pi}{\la^{\bo}}\Biggr )^{1/\ov N_c}+
m_Q\,\rm {Tr}\, \Pi \Biggr \}\,,\,\,\, \mu < \Lambda_{YM}\,.
\eq
Finally, at lower energies $\mu \ll \Lambda_{YM}$, after integrating out the last gluonium
field $S$ (with its mass scale $\sim \Lambda_{YM}$), one obtains the Lagrangian of pions:
\bq
L=\int \te \,\ote \, \Biggl \{ \rm {Tr}\,\sqrt {\Pi^\dagger \Pi}\Biggr \}\,\,+\nonumber
\eq
\bq
+\int \te \Biggl \{ -\nd  \Biggl (\frac{\det\,\Pi}{\la^{\bo}}\Biggr )^{1/\ov N_c}+
m_Q\rm {Tr}\, \Pi \Biggr \}\,,\,\,\, \mu\ll \Lambda_{YM}\,.
\eq
This describes weakly interacting pions with the smallest masses $m_{\pi}\sim m_Q$\,.
\footnote{\,
The vacuum value $\langle \Pi^i_{\ov j}\rangle=\cm^2\,\delta^i_{\ov j}$ recalls the scale
$\mu_C=\cm$ at which they were formed and so determines their "internal hardness", i.e. the
scale up to which they behave as pointlike particles.
}

So, this is the end of this story.
\footnote{\,
A short discussion of external anomalies (the 't Hooft triangles) is transferred to the appendix.
}

\section { Dual theory\,.\quad Definition\,}

\hspace {6 mm} The Lagrangian of the dual theory (at the scale $\mu \sim\la)$ is taken in the form
\cite{S1}\,:
\bq
{\ov L}=\int \te \,\ote \,\Biggr \{ {\rm Tr}\Biggl ( q^\dagger e^{\ov V} q + {\ov q}^\dagger
e^{-\ov V} {\ov q}\Biggr )+\frac{1}{(\mu^{\prime}_q)^2}{\rm Tr \Bigl (M^{\dagger}M\Bigr )}
\Biggr \}+
\eq
\bq
\int \te \,\Biggl \{ -\frac{2\pi}{\ov \alpha(\mu,\Lambda_q)}\, {\ov s}+\frac{1}{\mu_q}\,\rm {Tr}
\Bigl({\ov q}\,M\, q \Bigr ) +{\ov m}_Q(\mu)\,{\rm Tr \, M} \Biggr \}+{\rm h.c.}\,,\quad
  {\ov s}={\ov w}_{\alpha}^2/32\pi^2\,. \nonumber
\eq
Here\,:\, ${\ov a}(\mu)=\nd{\ov \alpha}(\mu)/2\pi$ is the running dual coupling (with its scale
parameter $\Lambda_q $),\, $a_f(\mu)=N_F f^2(\mu)/4\pi$  will be its running Yukawa coupling
(with its scale parameter $\Lambda_f$) with $f(\mu=\la)\sim\mu^{\prime}_q/\mu_q$\,,\, ${\ov w}_
{\alpha}$ is the dual gluon field strength. This theory has the exact $SU(\nd=N_F-N_c)$ gauge
symmetry, while in the chiral limit ${\ov m}_Q\ra 0$ the global symmetries are the same as in
the direct
theory. Under these symmetries the dual quarks and mesons $\rm M$ (mions) transform as:
\bq
q\,\,:\quad ({\ov N}_{F})_L^{\rm fl}\times (0)_R^{\rm fl}\times
(N_c/\nd)_B\times (N_c/N_F)_R\,,\nonumber
\eq
\bq
{\ov q}\,\,:\quad  (0)_L^{\rm fl}\times ( N_F)_R^{\rm fl}
\times (-N_c/\nd)_B \times (N_c/N_F)_R\,,
\eq
\bq
{\rm M}\,\,:\quad (N_{F})_L^{\rm fl}\times ({\ov N}_{F})_R^{\rm fl}\times
(1)_B\times (2\nd/N_F)_R\,.\nonumber
\eq

The mion fields ${\rm M}^i_{\ov j}$ in (4.1) are defined as pointlike ones. This is unlike the
pion fields ${\Pi}^i_{\ov j}$ of the direct theory, which appear as light pointlike fields only
at energies below the scale of chiral flavor symmetry
breaking, $\mu < \mu_C=\cm$. At higher scales $\mu \gg \cm$ they, strictly speaking, can't be
used at all (or, at best, can be resolved as composite fields of two current quarks).

To match  parameters of the direct and dual theories (see below), the normalizations at
$\mu=\la$ are taken as\,:
\bq
 \langle \,{\rm M}^i_{\ov j}\,\rangle_{\mu=\la} =\cm^2\,\delta^i_{\ov j},\quad
{\ov m}_Q(\mu=\la)=m_Q(\mu=\la)\equiv m_Q\,.
\eq
Besides, to match the values of gluino condensates, the
scale parameter $\Lambda_q$ has to be taken as \cite{IS}\,:
\bq
\Lambda_{\rm q} ^{\bd}=(-1)^{\nd}\,\Bigl (\mu_q^{N_F}/\la^{\bo}\Bigr )\quad \ra\quad
\langle S\rangle = \langle {-\,\ov s}\rangle\,, \quad \bd=(3\nd-N_F)\,.
\eq
\vspace{0.2cm}
\section { Dual theory with {\boldmath $\mu_q=\la $}\,. Conformal window}

\hspace {6 mm} With this choice, $|\Lambda_q|=\la$, see (4.4). In essence, this is the only natural
value for $\mu_q$, from a viewpoint of the direct theory. At $\mu_q\ll \la$ the value of $|\Lambda_q|$
will be either artificially small (at $N_F > 3N_c/2$), or artificially large (at $N_F < 3N_c/2$), see
(4.4). At $\Lambda_f\sim |\Lambda_q|=\la\,,\,(\,\mu^{\prime}_q \sim \mu_q\,)$\,, the dual theory
(which, self-consistently by itself, is considered to be in the UV-free logarithmic regime at $\mu
\gg \la$\,, with $a_f(\mu)<{\ov a}(\mu)$ at $\mu\gg \la$) enters, simultaneously with the direct
one, the superconformal regime at $\mu \sim \la$\,, with frozen couplings :
\,\,${\ov a}(\mu)\ra{\ov a}^{*}$ and $a_f(\mu)\ra a_f^{*}$. The dynamical dimensions
of chiral superfields are determined here by their R-charges, $\rm D=3|R|/2$\,,
so that, for instance, the distance dependence of the two-point
correlators $\langle \{{\ov Q}_{\ov j}Q^{i}(x)\}^{\dagger},\,{\ov Q}_{\ov l}Q^{k}(0)\rangle$ and
$\langle\{M^i_{\ov j}\}^{\dagger}(x),\,M^k_{\ov l}(0)\rangle$ is the same, etc. \cite{S1}.
Besides, all 't Hooft triangles are matched \cite{S1}. At present, no indication of possible
differences between the direct and dual theories is known in this perturbative superconformal
regime. So, let us go to lower energies where the physical scales originating from the chiral
symmetry breaking begin to reveal itself. What happens in the direct theory when reaching its
highest physical scale $\mu_H \sim \mu_C=\cm$ was described above in section 3.\\

In the dual theory and in the case considered, the highest physical scale $\mu_H$  is
determined by the constituent mass ${\ov \mu}_C$ of dual quarks, i.e. by the value of
their {\it coherent} condensate: $\mu_H={\ov \mu}_C=|\langle{\ov q}q\rangle| ^{1/2}_
{\mu=\la}= (m_Q\la)^{1/2}$, as this is parametrically larger in the conformal window
$3N_c/2 < N_F < 3N_c$ than the pole mass $m_q^{\rm pole}$ of dual quarks (\,$m_q(\mu)$ is the
running current mass of dual quarks, $m_q=m_q(\mu=\la)=\cm^2/\la\,,\,\gamma_q=\bd/N_F=
(3\nd-N_F)/N_F\,,\,\,\lym=\Bigl (\Lambda^{\bo}_Q \det m_Q \Bigr )^{1/3N_c}\,\,)$\,:
\bq
\frac{{\ov \mu}_C}{\la}=\Biggl (\frac{m_Q}{\la}\Biggr )^{1/2}\gg \frac{m_q^{\rm pole}}{\la}
\,,\quad \frac{m_q^{\rm pole}}{\la}=\frac{m_q(\mu=m_q^{\rm pole})}{\la}=\frac{\cm^2}
{\la^2}\Biggl (\frac{\la}{m_q^{\rm pole}} \Biggr )^{\gamma_q}=\frac{\lym}{\la}\,.\nonumber
\eq
This shows that, similarly to the direct theory, the dual theory is also in the same (dual)
DC - phase here, with appearance $N_F^2$ dual pions $\rm N_i^{\ov j}$ (nions) and
the large constituent masses ${\ov \mu}_C=(m_Q\la)^{1/2}$ of dual quarks when $\mu$
crosses the corresponding threshold region\,:\, ${\ov \mu}_2={\ov \mu}_C/(\rm several)\leq
\mu \leq{\ov \mu}_1=(\rm several)\, {\ov \mu}_C$\,. And similarly, all dual gluons also remain
massless at the same time. Therefore, the pattern of evolution through the threshold region is
universal, if either direct or dual theories are in the same DC - phase. So, using the
same reasonings as those described above in section 3 and making some simple substitutions of direct
parameters by dual ones, one obtains the effective dual Lagrangian at $\mu = {\ov \mu}_2$ in the form
(the meson and quark fields are normalized at $\mu=\la$ in (5.1)\,)\,:
\bq
{\ov L}=\int \te \,\ote \, \Biggl \{\frac{ z_M}{\la^2}\, {\rm Tr\,\Biggl ( M^{\dagger} M\Biggr )}
\,+ \rm {Tr\,\sqrt {N^\dagger N}}+{\ov Z}_2{\rm Tr}\Biggl( q^\dagger e^{\ov V} q+ {\ov q}^\dagger
e^{-\ov V} {\ov q}\Biggr ) \Biggr \}+\nonumber
\eq
\bq
+\int \te \,\Biggl \{-\frac{2\pi}{{\ov \alpha}(\ov{\mu}_2)}\,{\ov s}\, +W_q \Biggr \}\,,
\quad W_q=\frac{1}{\la}{\rm Tr \Biggl ( M N \Biggr )}+ m_Q\rm {Tr\, M}\,+\nonumber
\eq
\bq
+\Biggl ( \frac{\det\,{\rm N}}{\Lambda_{\rm q} ^{\bd} }\Biggr )^{1/N_c}
\Biggl [ \rm {Tr \Bigl ({\ov q} \,N}^{\,-1}\, q\Bigr )-N_F \Biggr ]\,,\quad
\quad {\ov Z}_2=\Biggl(\frac{\ov \mu_C}{\la}\Biggr )^{\bd/N_c}=
\Biggl (\frac{m_q}{{\ov \mu}_C} \Biggr )\,,
\eq
\bq
\langle {\rm M}^i_{\ov j}\rangle=\cm^2\,\delta^i_{\ov j}\,,\quad
\langle {\rm N}_i^{\ov j}\rangle=\langle {\ov q}^{\ov j}q_{i}
\rangle=-{\ov \mu}_C^2\,\delta_i^{\ov j}=- m_Q\la\,\delta_i^{\ov j}\,,\quad
m_q= \cm^2/\la\,.\nonumber
\eq

The factor $z_M\equiv z_M(\la\,, {\ov \mu}_1)\gg 1$ in (5.1) is the standard perturbative
renormalization factor of mion fields $\rm M$ in the interval ${\ov \mu}_1<\mu<\la$ (the fields
$\rm M\,, N$ and the dual quarks are frozen and do not evolve any more at $\mu<{\ov \mu}_2$
\,; besides, like the gluon fields, the mion fields $\rm M$ have no non-standard evolution
in the threshold region; and finally, here and everywhere below we neglect, as in sections 2 and
3, the dependence of the renormalization factors $z_M$ and ${\ov Z}_2$ on the quantum mion and nion
fields $m/\cm$ and $n/{\ov \mu}_C$\,, as this will influence the particle
mass values by non-parametric factors $\sim 1$ only, see also the footnote 5)\,:
\bq
z_M\equiv z_M(\la\,, {\ov \mu}_1)=\Bigl (\frac{\ov\mu_1}{\la} \Bigr )^{\gamma_M}=
1/z_q^2\,,\quad z_q\equiv z_q(\la\,,{\ov \mu}_1)=\Biggl
(\frac{{\ov \mu}_1}{\la} \Biggr )^{\bd/N_F}\ll 1\,,
\eq
where $z_q$ is the renormalization factor of the massless dual quarks due to the standard
perturbative evolution from $\mu=\la$ down to ${\ov \mu}_1=(\rm several)\,{\ov \mu}_C$\,.

And analogously to the direct theory, the factor ${\ov Z}_2$ in (5.1) is the overall
renormalization factor of the dual quark due to its evolution from $\mu=\la$ down to $\mu=
{\ov \mu}_2={\ov \mu}_C/(\rm several)$\,. It can be written in the form\,: ${\ov Z}_2=z_q\,
{\ov Z}_q$\,, where $z_q$ is due to the standard perturbative evolution in the interval ${\ov
\mu}_1<\mu<\la$\,, while ${\ov Z}_q$ is due to the additional non-standard evolution in the
threshold region ${\ov \mu}_2={\ov \mu}_C/(\rm several)<\mu<{\ov \mu}_1=(\rm several)\,{\ov \mu}_C$\,.

The heavy constituent dual quarks decouple at $\mu<{\ov \mu}_2$, and there remain the mions $\rm M$
and nions $\rm N$ and the pure gauge $SU(\nd)$ dual theory. As for its inverse coupling
$1/{\ov \alpha}(\mu)$\,, one obtains, similarly to the direct theory, that it increases from its
frozen value $1/{\ov \alpha}^{*}=O(1)$ at $\mu={\ov \mu}_1$ to a logarithmically large value at
$\mu={\ov \mu}_2$, due to the additional large renormalization factor ${\ov Z}_q$ of constituent dual
quarks. The whole evolution from $\mu=|\Lambda_q|$ down to $\mu<{\ov \mu}_2$ results in\,:
\bq
\frac{2\pi}{{\ov \alpha} (\mu<{\ov \mu}_2,{\ov \Lambda}_L)}=\Biggl \{ 3\nd \ln \frac
{\mu}{\Lambda_q}+\nd \ln \frac{1}{{\ov g}^{2}(\mu,\langle{\ov \Lambda}_L\rangle)}\Biggr \}-\Biggl \{
\ln \Biggl (\,\frac {\det \, \Bigl ({\ov \mu}_C \Bigr )^i_{\ov j}}
{\Lambda^{N_F}_q}\,\Biggr )-N_F\ln \frac{1}{{\ov Z}_2}\Biggr \}\,,\nonumber
\eq
\bq
\Bigl ({\ov \mu}_C \Bigr )^i_{\ov j}=\frac{1}{{\ov Z}_2}\Biggl ( \frac{\det\,{\rm N}}{\Lambda_{\rm
q}^{\bd} }\Biggr )^{1/N_c}\Biggl ({\rm N}^{\,-1} \Biggr )^i_{\ov j}\,\,\,\,,
\eq
where $\Bigl ({\ov \mu}_C \Bigr )^i_{\ov j}$ is the constituent mass of dual quarks, see (5.1).

Therefore, one obtains from (5.3) that the scale parameter ${\ov \Lambda}_L$ of
${\ov \alpha}(\mu,{\ov \Lambda}_L)$ is\,:
\bq
{\ov \Lambda}_L=\Biggl ( \frac{\det\,{\rm N}}{\Lambda_{\rm q}^{\bd} }\Biggr )^{1/3N_c}\,,\quad
|\langle {\ov \Lambda}_L\rangle |=\lym\,.
\eq

Lowering the scale down to $\mu<\lym$ and integrating out all gauge degrees of freedom through
the VY-procedure, one obtains the lowest energy Lagrangian\,:
\bq
{\ov L}=\int \te \,\ote \, \Biggl \{ \frac{z_M}{\la^2}\,{\rm Tr\,\Biggl ( M^{\dagger} M\Biggr )}
\,+ \rm {Tr\,\sqrt {N^\dagger N}}\Biggr \}\,\,+
\eq
\bq
+\int \te \Biggl \{\frac{1}{\la}{\rm Tr \Biggl ( M N \Biggr )}+ (\nd-N_F)
\Biggl ( \frac{\det\,{\rm N}}{\Lambda_{\rm q}^{\bd} }\Biggr )^{1/N_c}+
m_Q\rm {Tr\, M}\,\Biggr \}\,.\nonumber
\eq
Substituting $\Lambda_q$ from (4.4) and changing $\rm N\ra (-N)$\,, its superpotential can be
rewritten in a more convenient form \,:
\bq
W=\frac{1}{\la}{\rm Tr \Biggl (- M N \Biggr )}+ N_c
\Biggl ( \frac{\det\,{\rm N}}{\Lambda_{\rm Q}^{\bd} }\Biggr )^{1/N_c}+
m_Q\rm {Tr\, M}\,.
\eq

Therefore, the masses of mions $\rm M$ and nions $\rm N$  are, see (5.2)\,:
\bq
\mu_{M}\sim \mu_{N}\sim \Biggl (\frac{{\ov \mu}_C^2}{z_M} \Biggr )^{1/2}\sim\Biggl (\frac{m_Q
\la}{z_M} \Biggr )^{1/2}=\la \Bigl ( \frac{m_Q}{\la}\Bigr )^{3\nd/2N_F}\ll \lym\,.
\eq

On the whole, the mass spectrum looks here as follows\,: a)\, there is a large number of
hadrons made of non-relativistic (and weakly confined, the string tension is $\sqrt\sigma
\sim \lym\ll {\ov \mu}_C$\,) dual quarks, with their dynamical constituent masses ${\ov
\mu}_C=(m_Q\la)^{1/2}\ll \la$\,,\,\, b) there is a large number of gluonia with their
universal mass scale $\sim \lym$\,,\,\, c) the lightest are $\rm N_F^2$ mions $\rm M$ and
$\rm N_F^2$ dual pions $\rm N$ (nions) with masses $\mu_{M}\sim \mu_{N}\sim \la \Bigl
(m_Q/\la\Bigr )^{3\nd/2N_F}\ll\lym$ \,.

Comparing the mass spectra of the direct and dual theories, it is seen that they are very
different.\\

\section { Dual theory with {\boldmath $\mu_q=\cm $}\,.\quad
Conformal window}

\hspace {6 mm} Let us consider now this choice of parameters in (4.1). As will be shown below, this
choice will result in a much more close similarity of the mass spectra of direct and dual theories.

But first, one obtains in this case from (4.4): $|\Lambda_q|=(\cm^{N_F}/\la^{\bo})^{1/\bd}\ll\la$,
i.e. the scale parameter of the dual gauge coupling ${\ov \alpha}(\mu,\Lambda_q)$ is parametrically
smaller than those of the direct one. Moreover, it is parametrically smaller than even $\cm\,
: (|\Lambda_q|/\cm)=(\cm/\la)^{\bo/\bd}\ll 1$. But this means that these two theories are
clearly distinct in the perturbative interval $\cm <\mu < \la$. Indeed, the direct theory
entered already at $\mu<\la$ into the perturbative conformal regime, so that its coupling is
frozen at the value $\alpha^{*}$ and does not run.

As for the dual theory, the most natural boundary condition at $\mu=\la$ is to take the scale
factor $\Lambda_f$ of the Yukawa coupling $\Lambda_f\sim \Lambda_q$\,, this allows to consider
self-consistently the dual theory as UV-free by itself (but nothing will change essentially at
$\mu<\la$ also with $\Lambda_f\sim \la$\,, the Yukawa coupling will be $O(1)$ at $\mu\sim \la$ and
will decrease then logarithmically with decreasing $\mu<\la$\,, the problem will be that the Yukawa 
coupling will grow with increasing $\mu$ at $\mu>\la$).
With this choice, $a_f^{-1}(\mu=\la)=2\pi/{N_F\alpha_f(\mu=\la)}\sim({\ov
a})^{-1}(\mu=\la)=2\pi/{\nd{\ov \alpha}(\mu=\la)} \simeq\bd\ln(\la/\Lambda_q)\gg 1$\,.  Then,
with decreasing $\mu<\la$\,, both couplings of the dual theory increase
logarithmically but still remain $\ll 1$  at $|\Lambda_q|\ll\cm<\mu<\la$\,. So, the dual
theory will be in the weak coupling logarithmic regime at $\cm\ll\mu\ll\la$\,. Therefore, while
correlators of the direct theory behave already in a power-like fashion, those of the dual
one acquire only slow varying logarithmic renormalization factors.
\footnote{\,
Really, with so small value of $|\Lambda_q|\ll \cm$\,, the dual theory never enters the
conformal regime, see below.
}
Unfortunately, this is a price for a better similarity of both theories at lower scales
$\mu < \cm$.
\footnote{\,
From now on, to simplify all expressions, in all those cases when the dual theory is in the
weak coupling perturbative logarithmic regime, we will ignore the logarithmic renormalization
factors $z_q$ and $z_M$\, in calculations of mass spectra. In any case, because these
non-leading effects from $z_q\neq 1$ and $z_M\neq 1$ are only logarithmic, taking them into
account will not violate any power hierarchies and, besides, they are not of great importance
for numerical values of masses.
}

The current mass of dual quarks is now $m_q=\cm$, and it is much larger than the scale of their
condensate: $|\langle q\,{\ov q}\rangle | ^{1/2}=(m_Q\,\cm)^{1/2}$. So, they can't be now in
the collective coherent condensate phase, as their quantum fields are short ranged and will
fluctuate independently locally. Therefore, they can be treated simply as heavy
quarks (as their mass $\cm$ is much larger also than $|\Lambda_q|$).
\footnote{\,
Their non-zero vacuum condensate is now a pure quantum effect induced by the one-loop triangle
diagram\,: $\langle \,{\ov q}q(\mu=\cm)\,\rangle=\langle \,{\ov s}\,\rangle/\cm$\,, where
$\langle \,{\ov s}\,\rangle$ is the vacuum condensate of dual gluinos and $\cm\gg \lym$ is the
large current mass of dual quarks. This realizes the Konishi anomaly.
}
Going to lower scales $\mu \ll \cm$, they can be integrated out directly as heavy particles.
\footnote{\,
Because the dual quarks are confined, this leaves behind a large number of mesons and baryons
(with the mass scale $\sim\cm$\,, the string tension is $\sqrt \sigma\sim \lym\ll \cm$) made of
weakly interacting non-relativistic heavy dual quarks with the current masses $\cm$.
}

What remains then, is the $SU(\nd)$\, Yang-Mills theory (plus the mions ${\rm M}$) with the scale
parameter $\ov\Lambda_L$ of its coupling ${\ov \alpha}(\mu)$\,:
\bq
\frac{2\pi}{{\ov \alpha}(\mu,\ov\Lambda_L)}=3\nd\ln \frac{\mu}{\ov\Lambda_L}+\nd\ln
\frac{1}{{\ov g}^{2}(\mu,\langle\ov\Lambda_L\rangle)}\,,\quad -{\ov\Lambda}^3_L=\rm (\det M/\la^{\bo}
)^{1/\nd},\,\,\, |\langle {\ov\Lambda}_L\rangle|=\lym\,.
\eq
Therefore, at ${\Lambda}_{YM}\ll \mu \ll \cm$\,, the effective dual Lagrangian takes the form
(see the footnote 11)\,:
\bq
{\ov L}=\int \te \,\ote \,\Biggr \{ \frac{1}{\cm^2}\,{\rm Tr}\,\Bigl ({\rm M^{\dagger}M}\Bigr )
\Biggr \}+
\int \te \,\Biggl \{ -\frac{2\pi}{{\ov \alpha}(\mu,{\ov\Lambda}_L)}\,{\ov s}+m_Q {\rm Tr M} \Biggr \},
\eq

Finally, at  scales $\mu < \Lambda_{YM}$, using the VY-procedure
for integrating dual gluons, one obtains the lowest energy Lagrangian of mions\,:
\bq
{\ov L}=\int \te \,\ote \,\Biggr \{ \frac{1}{\cm^2}\,{\rm Tr}\,\Bigl ({\rm M^{\dagger}M}\Bigr )
\Biggr \}+ \nonumber
\eq
\bq
+\int \te \,\Biggl \{ -\nd \, \Biggl ( \frac {\det\, {\rm M}}{\la^{\bo}} \Biggr )^{1/\nd}
+m_Q {\rm Tr  M} \Biggr \}\,,\,\,\, \mu \ll \Lambda_{YM}\,\,.
\eq

This describes the mions ${\rm M}$ with masses $\sim m_Q$, interacting weakly through the
standard  superpotential.\\

On the whole, let us compare the direct and dual theories in the case considered.\,-

a) As was pointed out above, they are clearly different in the region $\cm < \mu < \la$.

b) There is a large number of colorless  hadrons, the heavy mesons (quasi-stable, decaying
into light pions or mions) and baryons (at least, those of lowest mass are stable) in both
theories, made of heavy non-relativistic (and weakly confined, the string tension is
$\sqrt \sigma\sim \lym\ll \cm$\,) constituents. In the direct theory these are the
constituent quarks with the dynamically generated masses $\mu_C=\cm$\,, while in the dual
theory these are simply the dual quarks itself with the same (but now current) masses $\cm$.
It seems that mesons are indistinguishable in both theories, but baryons are different
because they know about a number of colors and their masses will be different: $M_{\rm baryon}
=N_c\,\cm \neq {\ov M}_{\rm baryon}=\nd \,\cm$.

c) The remaining light particles in both theories at $\Lambda_{YM}\ll \mu \ll \cm$ are the
gauge ones, with respectively $N_c$ and $\nd$ colors, and pions (or mions). It is
important that both Yang-Mills theories, direct and dual, are at weak couplings in this interval of
scales, but have different numbers of colors. So, they are clearly different here.
\footnote{\,
Let us consider, for instance, the two-point correlators of the energy-momentum tensors in both
theories. Because both gauge couplings are small and contributions from pion or mion
interactions are already power suppressed at $\mu \ll \cm $\,, these correlators are dominated
by the lowest-order one-loop diagrams. The contributions of pions and mions are the same, but
contributions of gauge particles are different, as $N_c^2\neq \nd^{\,2}$.
}

d) There is a large number of (strongly coupled, quasi-stable due to decays into pions or mions)
gluonia in both theories, all with masses determined the same scale $\Lambda_{YM}$. So,
it seems, they look indistinguishable.

e) Finally, there are $\rm N_F^2$ of light pions (mions) with masses $\sim m_Q$ in both theories,
weakly interacting at low energies $\mu \ll \Lambda_{YM}$ through the same universal chiral
superpotential. Nevertheless, as it is, the interactions of pions and mions with
gluons at $\Lambda_{YM}\ll \mu\ll \cm$ are different in (3.10) and (6.2).\\

On the whole, it is seen that (with the logarithmic accuracy, see the footnote 11) the mass
spectra look very similar in both theories in this case (but not completely). But in many other
respects (see above) the direct and dual theories are clearly different.

\section { {\boldmath $N_c < N_F < 3N_c/2$}\,}

\hspace {6 mm} There are two possible ways to interpret the meaning of the Seiberg dual theories at
$N_c < N_F < 3N_c/2$.\,\,\,-

{\bf a)}\, The first variant is similar to those which is the only possibility in the conformal
window $3N_c/2 < N_F < 3N_c$\,. I.e., the description of all light degrees of freedom of
the direct theory in terms of massless quarks $Q,\,{\ov Q}$ and gluons remains adequate in the
interval of scales $ \mu_H\ll \mu \leq \Lambda_Q$, where $\mu_H\ll \la$ is the highest physical
mass scale due to $m_Q\neq 0$, and there are no massive particles with masses $\sim \Lambda_Q$
in the spectrum at $m_Q \ll \la $. In comparison with the conformal behavior, the difference
is not qualitative but only quantitative: the strong coupling does not approach the constant
value $\alpha^{*}$ at $\mu \ll \Lambda_Q$ but continues to grow. Nevertheless, the
non-perturbative contributions are power suppressed until $\mu \gg \mu_H\,$, and one obtains
the right answers for all Green functions by resummation of standard perturbative series with
massless quarks and gluons. The dual theory is interpreted then as a possible alternative but
equivalent (weak coupling) description. This variant can be thought of as some formal
'algebraic duality', i.e. something like 'the generalized change of variables'.

{\bf b)}\,\, The second variant is qualitatively different (it is sometimes referred to as
'confinement without chiral symmetry breaking', i.e. due only to $\la\neq 0$ at $m_Q\ra 0$).
It implies that, unlike the variant 'a', the non-perturbative contributions become essential
already at $\mu\sim \la$, resulting in a high scale confinement with the string tension
$\sqrt \sigma\sim \la$ which binds direct quarks and gluons into colorless hadron states
with masses $\sim \la$. This can be thought of, for instance, as follows. At $N_F$ close to
$3N_c$ the value of $a^{*}=N_c\alpha^{*}/2\pi$ is small. As $N_F$
decreases, $a^{*}$ increases and becomes $\simeq 1$ at $N_F$ close to $3N_c/2$.  When
$N_F<3N_c/2$, the coupling $a(\mu)$ exceeds some critical value $a^{(\rm crit)}=O(1)$ already at
$\mu\sim \la$ and it is assumed that, for this reason, the theory is now in another phase.
The strong non-perturbative confining gauge interactions begin to operate at the scale $\sim \la$\,,
resulting in appearance of large number of colorless hadrons with masses $\sim \la$.
So, the use of old massless quark and gluon fields for description of light
degrees of freedom at $\mu \ll \la$ becomes completely inadequate.
\footnote{\,
This is especially visible at $N_F=N_c+1$ where, for instance, the gauge degrees of freedom
are not present at all amongst light ones in the dual theory.
}

Instead, the new (special solitonic ?) light degrees of freedom are formed at the
scale $\sim \la$ as a result of these strong non-perturbative effects. These are the dual quarks
and gluons and dual mesons $\rm M$ (mions), with their sizes $\sim 1/\la$ and the internal
hardness scale $\sim \la$ (i.e. they appear as point-like at $\mu < \la$). These new light
particles are described by fields of the dual theory. So, this variant 'b' can be thought of
as 'the physical duality', in a sense that the dual theory is really the low energy
description of the original theory at $\mu<\la$\,.\\

Now, we would like to present arguments against the variant 'b'. The above described scenario
of 'confinement without chiral symmetry breaking' implies that, even at $m_Q\ra 0$\,,
there will be a large number of massive (with masses $\sim\la$) colorless hadrons $H_n$ in
the spectrum, both non-chiral made of $(Q^{\dagger}\,,Q)$ or $({\ov Q}^{\dagger}\,,{\ov Q})$
quarks, and chiral made of $({\ov Q}\,,Q)$ quarks, etc.

Let us consider, for instance, the action of the simplest colorless chiral superfield $\p$\,\,
\footnote{\,
Or any other colorless spin zero or higher spin chiral superfield composed in some way from
$Q^i,\,{\ov Q}_{\ov j}$ and the gauge field strength $W_{\alpha}$, for
instance\, $({\ov Q}_{\ov j}T^a Q^i)\,W^a_{\alpha}$\,,  etc.
}
on the vacuum state: $\p\,|0\rangle$. This operator will excite from the vacuum not only,
say, the massless one-mion state $|M^i_{\ov j}\rangle$, but also
many one-particle states of massive chiral hadrons $|\Psi_n\rangle$.
Let $\Psi_{\ov j}^i$ be the regular chiral superfield of anyone of such hadrons. Then in the
effective Lagrangian describing theory at the scale $\mu\sim \la$ there should be a term in
the superpotential which describes the nonzero mass $\sim \la$ of this chiral hadron.
But the standard regular term $\la \rm {Tr}(\Psi\,\Psi)$ is not allowed as it breaks
explicitly the chiral flavor $SU(N_F)_L\times SU(N_F)_R\,$ symmetry (and R-charge), and it
seems impossible to write in the superpotential at $m_Q\ra 0$ the appropriate regular mass terms
for massive chiral hadron superfields with masses $\sim \la$.
\footnote{\,
One can try to 'improve' situation  multiplying the regular chiral superfield $\Psi_{\ov j}
^i$ by the chiral superfields $(\p/\la^2)^{-1}$ and $(\det\,\p /\la^{2N_F})^{1/\Delta}$ to
build up the term in the superpotential with appropriate quantum numbers, but all such
terms are\, singular at $\langle 0|\p |0 \rangle \ra 0$\,, and so all this will not result in
obtaining the genuine regular mass term for this hadron. Trying to use the dual quark fields
$q$ and ${\ov q}$ together with $\Psi$ also does not help as $\langle{\ov q}q\rangle\ra 0$
at $m_Q\ra 0$.

One can also consider the variant of 'b' when the direct color is not confined.\\
From our point of view, this is the only realistic variant in the chiral limit $m_Q= 0$\,.
Because, at least in SQCD, the strong coupling $a(\mu\sim \la)\gtrsim 1$ does
not mean really that the scale of confining forces is $\sim \la$ (in other words, that the
string tension is $\sqrt \sigma\sim \la$). The underlying reason is that the role of the
order parameter for the confinement plays not $\la$ by itself, but rather the scale of the
gluino condensate, i.e. $\sqrt \sigma\sim\lym= \langle \lambda\lambda\rangle^{1/3}$. But
$\langle \lambda\lambda\rangle \ra 0$ at $m_Q\ra 0$\,. So, there will be no confinement at
all in the chiral limit $m_Q= 0$\,, and the regimes at $m_Q=0$ and $N_c<N_F<3N_c$ can be
called more adequately as 'the pure perturbative massless regimes with neither confinement, nor
chiral symmetry breaking', down to $\mu\ra 0$. They are : conformal at $3N_c/2<N_F<3N_c$\,, and
strong coupling at $N_c<N_F<3N_c/2$\,, see (7.4) below).

Then, in the variant 'b', the absence of confinement at $m_Q=0$ implies that the individual quarks
$Q^i$ and ${\ov Q}_{\ov j}$ will be present in the spectrum and they will be massive,
with masses $\sim \la$ (because there are no such light fields in the dual theory). And one
will face the same problem that it is impossible
to write in the superpotential the right regular mass term for these quarks.
}

In other words, the appearance in the spectrum of massive chiral flavored (and R - charged)
particles with masses $\sim \la$ at $m_Q\ra 0$ seems impossible without the spontaneous
breaking of $SU(N_F)_L\times SU(N_F)_R\,$ (and R-charge) symmetry.

If symmetry is broken spontaneously, there should be then the appropriate non-invariant
(elementary or composite) chiral superfield(s) $\phi_k$ which condenses in the vacuum with the
large value: $\langle 0|\phi_k|0\rangle =\phi_k^{(o)} \sim \la$. This condensate can give then,
in principle, the masses $\sim \la$ to chiral hadron superfields. But this basic condensate
$\phi_k^{(o)}$ should figure then explicitly in the low energy Lagrangian, from which its
numerical value in a chosen vacuum should be determined. The dual theory claims that it gives
a right description at low energies. But no one so large chiral vacuum condensate $\phi_k^
{(o)}\sim\la$ appear neither in the dual theory, nor in the direct one. We conclude that,
indeed, the chiral flavor $SU(N_F)_L\times SU(N_F)_R\,$ and R-charge symmetries are not
broken spontaneously at $m_Q\ra 0$.

So, the above considerations imply that the scenario 'b' is incompatible with unbroken
$SU(N_F)_L\times SU(N_F)_R\,$ (and R-charge) symmetries at $m_Q/\la\ra 0$.\\

Therefore, we will consider below the scenario 'a' only in which the nonzero particle masses
arise only due to breaking of the $SU(N_F)_L\times SU(N_F)_R\,$ and R-charge symmetries due to
$m_Q\neq 0$, and these masses are all much smaller than $\la$ at $m_Q\ll \la$. Because in this
variant the spectrum of light (i.e. with masses $\ll \la$) particles is known in both theories,
direct and dual, it becomes possible, in addition to the 't Hooft triangles, to compare also the
values of some special correlators in the perturbative range of energies where all particles
can still be considered as being massless ($\mu_H\ll\mu\ll \la$, where $\mu_H$ is the highest
physical scale due to $m_Q\neq 0$). These are the two-point correlators of external conserved
currents, say, the baryon and $SU(N_F)$ flavor currents, as these can be
computed in the perturbation theory even in the strong coupling region. Really, it is more
convenient to couple these conserved currents with the external vector fields and to consider
the corresponding external $\beta_{\rm ext}$\,-functions. Such $\beta_{\rm ext}$-functions have the
form (see e.g. \cite{KSV}) :
\bq
\frac{d}{d\,\ln \mu}\,\frac{2\pi}{\alpha_{ext}}= \sum_i T_i\,\bigl (1+\gamma_i
\bigr )\,,
\eq
where the sum runs over all fields which can be considered as being massless at a given scale
$\mu$, the unity in the brackets is due to one-loop contributions while the anomalous
dimensions $\gamma_i$  of fields represent all higher-loop effects.

So, let us equate the values of such $\beta_{\rm ext}$-functions in the direct and dual theories
at scales $\mu_H\ll\mu \ll \la$. The light particles in the direct theory are the original
quarks $Q,\,{\ov Q}$ and gluons, while in the dual theory these are the dual quarks $q,\,
{\ov q}$ and dual gluons, and the mions ${\rm M}$. For the baryon currents one obtains:
\bq
N_F N_c\,\Bigl ( B_Q=1 \Bigr )^2\,(1+\gamma_Q)=N_F \nd \,\Bigl ( B_q=\frac{N_c}{\nd}
\Bigr )^2\,(1+\gd)\,,
\eq
while for the $SU(N_F)_L$ (or $SU(N_F)_R\,)$ flavor currents one obtains:
\bq
N_c\,(1+\gamma_Q)=\nd \,(1+\gamma_q)+N_F\,(1+\gamma_M)\,.
\eq
Here, the left-hand sides are from the direct theory while the right-hand sides are from the
dual one, $\gamma_Q$ is the anomalous dimension of the quark $Q$\,, while $\gamma_q$ and
$\gamma_M$ are the anomalous dimensions of the dual quark $q$ and the mion ${\rm M}$.

Now, at $\mu_{H}\ll\mu\ll \la$ the dual theory is IR-free and both its couplings are small in this
range of energies, ${\ov a}(\mu)\ll 1\,,\, a_f(\mu)\ll 1$\,. So, $\gamma_q(\mu)\ll 1$ and $\gamma
_M(\mu)\ll 1$ are both also logarithmically small at $\mu\ll\la$\,. It is seen then that \,(7.2)
and (7.3) are incompatible with each other as they predict different values for the infrared
limit of $\gamma_Q$. We conclude that both correlators can not be equal
simultaneously in the direct and dual theories, and so these two theories are different.
\footnote{\,\,
Taking the IR value $\gamma_Q\ra (N_c/\nd-1)=(2N_c-N_F)/(N_F-N_c)$ from (7.2) as a concrete
example, and using the perturbative NSVZ $\beta$-function \cite{NSVZ}, one obtains
the perturbative IR-behavior of the strong coupling $\alpha(\mu)$\,:\\
$$\frac{d a(\mu)}{d\ln \mu}\equiv \beta(a)=-\, \frac{a^2}{1-a}\, \frac{\bo-N_F\gamma_Q}
{N_c}\,\,,\nonumber$$
$$a(\mu)\equiv \frac{N_c\alpha(\mu)}{2\pi}, \quad \bo=(3N_c-N_F), \quad
\gamma_Q\equiv \frac{d\ln z_Q}{d \ln \mu}\,,\quad z_Q(\la,\mu)=\Bigl (\frac{\mu}{\la}
\Bigr )^{\gamma_Q}\ll 1\,,$$
\bq
\gamma_Q=\frac{(2N_c-N_F)}{(N_F-N_c)}\,,\quad a(\mu)= \Bigl (\frac{\Lambda_Q}{\mu}\Bigr )^
{\nu}\gg 1\,,\quad \nu=\frac{3N_c-2N_F}{N_F-N_c}\,,\quad (\mu/\Lambda_Q)\ll 1\,.
\eq

In this case, the behavior of $a(\mu/\la)$ looks as follows. As $z=\mu/
\Lambda_Q$ decreases from large values, $a(z)$ increases first in a standard
way $\sim (1/\ln z)$. At $z=z_o\sim 1\,\,\, a(z)$ crosses unity. At this point
$\gamma_Q$ crosses the value $b_o/N_F=(3N_c-N_f)/N_F$. As a result, the $\beta$-function  is
smooth, it has neither pole nor zero at this point and remains negative all the way from the
UV region $z\gg 1$ to the IR region $z\ll 1$, while $a(z)$ grows in the
infrared region in a power-like fashion, see (7.4). On the other
hand, it is not difficult to see that the IR-value of $\gamma_Q$ obtained from (7.3) with
$\gamma_q\ra 0,\,\gamma_M\ra 0$ is incompatible with the NSVZ\, $\beta$-function.
}
\\

Nevertheless, it is of interest to compare their mass spectra which will reveal itself at
lower energies.

As for the direct theory, as was argued above, its qualitative properties don't differ much
from those described before for the conformal window. The main quantitative difference is
that the gauge coupling $\alpha(\mu)$ does not freeze at $\mu\ll \la$ but continues to
grow (for instance, as in (7.4)), until $\mu$ reaches the scale of the dynamical chiral
symmetry breaking,\, $\mu\sim \mu_C=\cm$\,. But after crossing the threshold region $\mu_2=
\cm/(\rm several)<\mu<\mu_1=(\rm several)\, \cm$ the coupling also becomes logarithmically
small, and the effective Lagrangian has the same form as in (3.2).  Indeed, as was
emphasized in section 3, this is independent of the explicit form of the quark perturbative
renormalization factor $z_Q(\la,\,\mu_1)$\, which
enters the evolution of the coupling $\alpha^{-1}(\mu)$ in the region $\mu_1 <\mu <\la$\,,
because this last cancels in (3.8) independently of its explicit form. The only restriction
is that the dynamical scenario has to be self-consistent. I.e., the constituent mass $\mu_C$
of quarks has to be larger than their perturbative pole mass, $\mu_C=\cm > m_Q^{\rm pole}$\,,
so that it will stop the perturbative massless RG-evolution before this will be done by $m_Q^
{\rm pole}$\,. It is not difficult to check that this is fulfilled with $\gamma_Q=(2N_c-N_F)/
(N_F-N_c)$ from (7.2)\,:
\bq
\frac{m_Q^{\rm pole}}{\la}=\frac{m_Q}{\la}\Biggl (\frac{\la}{m_Q^{\rm pole}} \Biggr )^{\gamma
_Q}=\Biggl (\frac{m_Q}{\la}\Biggr )^{(N_F-N_c)/N_c}\ll \frac{\cm}{\la}=\Biggl (\frac{m_Q}{\la}
\Biggr )^{(N_F-N_c)/2N_c}\,.\nonumber
\eq

So, below the threshold region $\mu<\mu_2$\,, all equations and all qualitative properties
of the direct theory described above for the conformal
window remain the same also in the region $N_c < N_F < 3N_c/2$.\\

As for the dual theory, we also consider here two variants for the scale parameter $\mu_q$
in (4.1):\, a) $\mu_q =\la$\,,\, and b)\, $\mu_q = \cm$\,. \\

{\bf a)\,\,\,\boldmath $\mu^{\prime}_q \sim\mu_q=\la$\,.}\\

In this case the scale parameter $\Lambda_q$ of the dual gauge coupling ${\ov \alpha}(\mu)$ is\,
: $|\Lambda_q|\sim\Lambda_f\sim\la$, see (4.4), both couplings ${\ov a}(\mu)$ and $a_f
(\mu)$ are $\lesssim 1$ at $\mu=\la$ and both decrease logarithmically when $\mu$ is going down
from $\mu \sim \la$ to $\mu_H\sim \cm^2/\la\ll \la$.

In the case considered, the current mass of dual quarks is (see the footnote 11)\,:
\bq
m_q=\langle {\rm M}\rangle /\mu_q =
\cm^2/\la\,,\quad m_q \gg |\langle {\ov q}\,q\rangle| ^{1/2}= (m_Q\,\la)^{1/2}\,,
\eq
i.e. it is much larger than the scale of their condensate, so that the dual theory is here in
the same HQ (heavy quark) phase as it was in section 6. Therefore, at lower scales all quarks
can simply be integrated out as heavy (and weakly confined, the string tension
is $\sqrt \sigma \sim\lym \ll \cm^2/\la$\,) particles, leaving behind a large number of
hadrons with masses $\sim \cm^2/\la$ composed of non-relativistic dual quarks. After this,
one obtains the effective Lagrangian in the form:
\bq
{\ov L}=\Biggr \{ \frac{1}{\la^2}{\rm Tr\Bigl (M^{\dagger}M\Bigr )}
\Biggr \}_D+ \Biggl \{ -3\nd\,{\ov s}\Biggl [\ln \frac{\mu}{{\ov \Lambda}_L}
+\ln\frac{1}{{\ov g}^2(\mu/\lym)}\Biggr ]+m_Q {\rm Tr\,M }\Biggr \}_F\,,
\eq
\bq
{\ov \Lambda}_L^3=-(\det {\rm M}/\la^{\bo})^{1/\nd}\,,\quad |\langle {\ov \Lambda}_L\rangle|=
\Lambda_{YM}\,,\quad \Lambda_{YM}\ll \mu \ll \cm^2/\la\,.\nonumber
\eq

Going down in energy and integrating out all gluonia (with masses $\sim \Lambda_{YM})$
through the VY- procedure, one obtains finally:
\bq
\ov L=\Biggr \{ \frac{1}{\la^2}\,{\rm Tr\,\Bigl (M^{\dagger}M\Bigr )}
\Biggr \}_D+\Biggl \{ -\nd \, \Biggl ( \frac {\rm \det\, M}{\la^{\bo}} \Biggr )^{1/\nd}
+m_Q{\rm Tr  \,M} \Biggr \}_F\,,\,\,\, \mu \ll \Lambda_{YM}\,\,.
\eq
This describes the mions ${\rm M}$ with the masses:
\bq
\mu_M\sim m_Q\Biggl (\frac{\la^2}{\cm^2}\Biggr )\sim m_Q \Biggl (\frac{\la}{m_Q}\Biggr )^
{\nd/N_c}\,,\quad m_Q\ll \mu_M \ll \Lambda_{YM}\,,
\eq
interacting weakly through the standard superpotential.\\

So, comparing the mass spectra of the direct and dual theories one sees that they are very
different.\\

{\bf b)\,\,\,{\boldmath $\mu_q=\cm $\,.}}\\

With $\Lambda_f\sim |\Lambda_q|$\,, both dual scale factors become very large with this choice of
$\mu_q $\,, see (4.4):
\bq
|\Lambda_q|=\Biggl(\frac{\la^{\bo}}{\cm ^{N_F}}\Biggr)^{(- 1/\,{\ov {\bo}\,} )}\,\gg \la\,.
\eq

But we can ignore the high energy region $\mu>|\Lambda_q|$ where the dual theory is strongly
coupled, and to start directly with $\mu\lesssim\la \ll |\Lambda_q|$\,, where both couplings are
already logarithmically small : $2\pi/{\alpha_f}(\mu=\la)\sim 2\pi/{\ov \alpha}(\mu=\la)\simeq
\bd\ln (|\Lambda_q|/\la)\gg 1$\,, and both continue to decrease logarithmically with decreasing
$\mu$ at $\cm<\mu<|\la|$\,. The region
$\cm < \mu < \la$ was discussed above, see (7.2),(7.3). So, let us consider now $\mu < \cm$.

The regime in this case b) is qualitatively the same as in the case a) above  (see also
the footnote 11), i.e. there are now even heavier dual quarks with the current mass $m_q=\cm$
(and even smaller condensate), the intermediate mass gluonia and smallest mass mions M. And
(7.6),\,(7.7) remain essentially the same, only the factor $1/\la^2$ in the meson Kahler term
is replaced now by $1/\cm^2$. Due to this, the masses $\mu_M$ of mions are now:
\bq
m_q\sim \cm \gg M_{\rm gl}\sim \Lambda_{YM}\gg \mu_M\sim m_Q\,.
\eq

So, in this case the mass spectra of the direct and dual theories (with the logarithmic
accuracy) are much more similar, as it was in section 6 in the conformal window. But all
differences (at scales $\mu <\cm$) described in section 6 also remain.\\

{\bf c)\,\,\quad {\boldmath $N_F=N_c+1$\,.}}\\

As for the direct theory, this point is not special and all equations and results described
before remain without changes.
\footnote{\,
Really special is the point $N_F=N_c$\,, as $\cm=\la$ in this case, even in the chiral limit
$m_Q\ra 0$, see (2.7). We do not consider this case here.
}

But this point is somewhat special for the dual theory because
its field content consists in this case of light mesons $\rm M^i_{\ov j}$ and
baryons $\rm B_i\,,\,{\ov B}^{\ov j}$\, only\, \cite{S1}.

The dual Lagrangian at $\mu < \la$ is supposed to have the form\, \cite{S1}\, :
\bq
{\ov L}=\int \te \ote \,\Biggl \{{\rm \frac{M^{\dagger}M}{\mu^2_{M}}+\frac{B^{\dagger}B\,\,+
{\ov B}^{\dagger}{\ov B}}{\mu_{B}^{2(N_c-1)}}} \Biggr \}+\nonumber
\eq
\bq
+\int \te \,\,\Biggl \{\rm{\frac{ Tr ({\ov B}M B)-\det M}{\la^{\bo}}+m_Q{\rm Tr\, M}}
\Biggr \}\,,\quad \mu\ll \la\,.
\eq
Here\,: the scale factors $\mu_M$ and $\mu_B$ in the Kahler terms are due to non-canonical
dimensions of meson and baryon fields (\,$M\ra {\ov Q}Q,\,\,B\ra Q^{N_c}$\,).

As for the interval of energies above the highest physical scale $\mu_H\,,\,\, \mu_H \ll \mu
\ll \la$, the equations (7.2,\,7.3) still hold in this case, with substitution: $\nd=1,\,
\gamma_q\ra \gamma_B$, and $\gamma_M,\, \gamma_B\ra 0$. So, they remain incompatible.

At lower energies, the meson and baryon masses can be obtained directly from the Lagrangian
(7.11)\,:
\bq
M_{M}\sim \, m_Q\Bigl (\frac{\mu_M}{\cm}\Bigr )^2,\quad M_B=M_{\ov B}\,\sim \frac
{\cm^2\,\mu_B^{2(N_c-1)}}{\la^{\bo}}\,.
\eq
So, at $\mu_M\sim \mu_B\sim \la\,: M_{M}\sim \la (m_Q/\la)^{(N_c-1)/N_c}\,,\,B_B\sim \la (m_Q/\la)
^{1/N_C}$. Let us recall (see above) that the mass spectrum of the direct theory consists here also
of a large number of flavored hadrons with the mass scale $\sim \cm\sim \la (m_Q/\la)^{1/(2N_c)}$\,, a
large number of gluonia with masses $M_{\rm gl}\sim \Lambda_{YM}\sim \la(m_Q/\la)^{(N_c+1)/3N_c}$\,,
and $N_F^2$ light pions with masses $\sim m_Q$.\\

\section {$ N_F> 3N_c$\,}

\hspace {6 mm} For completeness, let us also consider this region.\\

As for the direct theory, it is IR-free in this region ($\bo < 0$) at $ m_Q^{\rm pole}<\mu<\la$\,.
So, in a sense, it is very "simple" at $\mu \gg \lym$ (but at the price that it is now, at best,
strongly coupled in the UV-region $\mu \gg \la$ and, at worst, can't be defined self-consistently in
UV by itself and needs the UV completion).

The current quark mass $m_Q=m_Q(\mu=\la)\ll \la$ is much larger now than the scale of its chiral
condensate $\cm\ll m_Q$\,, see (2.7), and this power hierarchy persists at lower energies as the
RG - evolution here is only logarithmic at $ \lym \ll\mu<\la$\,. Therefore, the direct theory is
at $N_F>3 N_c$ in the HQ (heavy quark) phase, so that there is a standard weak coupling slow
logarithmic evolution in the region $m_Q^{\rm pole}\ll \mu \ll \la ,\,\, m_Q^{\rm pole}\equiv
m_Q (\mu=m_Q^{\rm pole})=z_Q^{-1}(\la,\,\mu=m_Q^{\rm pole})\,m_Q \gg m_Q $, where $z_Q(\la,\,
\mu=m^{\rm pole}_Q)\ll 1$ is the standard perturbative logarithmic renormalization factor of
massless quarks, and the highest physical scale is now\,: $\mu_H=m_Q^{\rm pole} \gg \Lambda_
{\rm YM}\gg \cm$. At $\mu \ll m_Q^{\rm pole}$ all quarks can be integrated out as heavy
(and weakly confined, the string tension is $\sqrt \sigma\sim \lym\ll m^{\rm pole}_Q $\,, their
vacuum condensate $\langle{\ov Q}Q (\mu=m_Q^{\rm pole})\rangle=\langle S\rangle /m_Q^{\rm pole}$
is due to a simple quantum one-loop contribution~) non-relativistic particles,
leaving behind a large number of mesons and baryons made of these non-relativistic quarks,
with masses : ${\rm M_{meson}\sim m^{\rm pole}_Q,\,\, M_{baryon}\sim N_c \,m^{\rm pole}_Q}$\,.
Evidently, there are no additional lighter pions now.

Using (2.2,\,2.3) to match couplings at $\mu= m^{\rm pole}_Q$\,, one obtains at lower energies
$\mu < m^{\rm pole}_Q$ the Yang-Mills Lagrangian with the scale factor of its gauge coupling
$\Lambda_{YM}=(\la^{\bo}\,m_Q^{N_F})^{1/3N_c}\ll m_Q$\,, so that this Yang-Mills theory is in
the weak coupling regime at $\lym\ll \mu<m_Q^{\rm pole}$\,. Finally, it describes strongly
coupled gluonia with masses $M_{\rm gl}\sim \Lambda_{YM}\ll m_Q^{\rm pole}$, and these are the
lightest particles in this case. So, this is the end of this short story in the direct theory.\\

As for the dual theory, as before, its mass spectrum depends on the value of $\mu_q$.\\

{\bf a.\quad Dual theory with}{\boldmath \quad $\mu^{\prime}_q\sim \mu_q=\la$\,.}\\

$\Lambda_f\sim |\Lambda_q|\sim\la$\,, see (4.4), but there are no particles with masses $\sim
\la$, similarly as it was for the direct theory in section 7. The dual theory is taken as UV-free
in this case and it enters the strong coupling perturbative regime at $\mu_H<\mu<\la$. For
definiteness, let us use the values of the dual quark and mion anomalous dimensions from (7.2),
(7.3) with $\gamma_Q\ra 0$ at $\mu\ll \la$\,:
\bq
\gamma_q=\frac{\nd}{N_c}-1\,,\quad \gamma_M=-\frac{\nd}{N_c}\,.
\eq

Now, it is seen that the dynamical constituent mass of dual quarks ${\ov \mu}_C$\,
is parametrically larger here than their pole mass $m_q^{\rm pole}$\,:
\bq
{\ov \mu}_C=(m_Q\la)^{1/2}\gg m_q^{\rm pole}\,,\quad m_q^{\rm pole} =\frac{\cm^2}{\la}\Bigl
(\frac{\la}{m_q^{\rm pole}}\Bigr )^{\gamma_q}=\la\Bigl (\frac{\cm^2}{\la^2} \Bigr )^{1/(1+\gamma_q)}
=m_Q\,.\nonumber
\eq

So, $\mu_H={\ov \mu}_C$ and the dual quarks are in the (dual) DC - phase. The
Lagrangian has the same form (5.1), all equations (5.3-5.6) remain the same and, instead
of (5.7), the masses of mions and nions look now as\,:
\bq
\frac{\mu_M}{\la}\sim \frac{\mu_N}{\la}\sim \Biggl (\frac{{\ov \mu}^2_C}{z_M\la^2}\Biggr )^{1/2}
=\Biggl(\frac{m_Q}{\la}\Biggr )^{(N_F+N_c)/4N_c}\,,\quad z_M=\Bigl (\frac{{\ov \mu}_C}{\la}
\Bigr )^{\gamma_M}\gg 1\,.
\eq

On the whole, the mass spectrum of the dual theory includes in this case\,: a) a large number
of flavored hadrons with their mass scale $\sim {\ov \mu}_C$\,, made of dual quarks with
the constituent masses ${\ov \mu}_C=(m_Q\la)^{1/2}\ll \la$\,,\,\, b) ${\rm N}_F^2$ mions and
${\rm N}_F^2$ nions with masses $\mu_M\sim\mu_N\sim\la\Bigl (m_Q/\la\Bigr )^{(N_F+N_c)/4N_c}\ll
{\ov \mu}_C$\,,\, c) a large number of gluonia with the mass scale $\sim \lym\ll \mu_M\sim
\mu_N$\,.\\

{\bf b.\quad Dual theory with}{\boldmath \quad $\mu_q=\cm$\,.}\\

With the choice $|\Lambda_q|\sim \Lambda_f$\,, both are very small in this case, see (4.4). The dual
theory is also taken as UV-free at
$\mu>|\Lambda_q|$ in this case, and it will also enter the strong coupling perturbative regime
at $\mu_H<\mu<|\Lambda_q|$\,. The boundary conditions for the dual gauge coupling ${\ov a}=
\nd{\ov \alpha}/2\pi$ and the Yukawa coupling $a_f=N_F f^2/2\pi$ at $\mu=\la$ look as:
${\ov a}(\mu=\la)\sim a_f(\mu=\la)\sim 1/\ln(\la/|\Lambda_q|)\ll 1$\,. In the perturbative
regions:\,\,$\Lambda_{YM}\ll\mu\ll\la$ for the direct theory and $\lym\ll |\Lambda_q|
\ll \mu \ll \la$ for the dual one, both theories are now in the weak coupling logarithmic regime.
The direct theory - because it is IR-free at $|\Lambda_q|\ll m^{\rm pole}_Q \ll \mu \ll \la$,
while its coupling $a(\mu)$ increases logarithmically at $\lym\ll \mu\ll m^{\rm pole}_Q$
but is still small. The dual theory - because $|\Lambda_q|\sim \Lambda_f$ are so small, both
its couplings ${\ov a}(\mu)$ and $a_f(\mu)$ increase logarithmically with decreasing $\mu<\la$
but still remain small at $\mu\gg |\Lambda_q|$\,. So, at $m^{\rm pole}_Q \ll \mu \ll \la $ the
direct and dual theories are both in the weak coupling logarithmic perturbative massless regime,
the equations (7.2,\,7.3) can be used with all $\gamma_Q,\,\gamma_q,\,\gamma_M \ll 1 $ now, and
they are incompatible.

At $\mu\ll |\Lambda_q|$ the dual theory is in the strong coupling regime ${\ov a}(\mu)\gg 1\,,\,
a_f(\mu)\gg 1$\,, and we use for the anomalous dimensions $\gamma_q$ and $\gamma_M$ the values
(8.1).

The hierarchies in the dual theory at $\mu=\la\gg |\Lambda_q|$ look as\,:
\bq
m_q=\cm\ll {\ov \mu}_C=|\langle{\ov q}q \rangle |^{1/2}=(m_Q\cm)^{1/2}\ll |\Lambda_q|\,,\quad
\frac{|\Lambda_q|}{\la}=\Bigl ( \frac{\cm}{\la}\Bigr )^{N_F/\bd}\ll 1\,,\nonumber
\eq
where $m_q$ is the current quark mass and ${\ov \mu}_C$ is its (possible) constituent mass. The
evolution in the interval $|\Lambda_q|<\mu<\la$ is only logarithmic ( all logarithmic effects are
neglected in what follows) and the hierarchies at $\mu\sim |\Lambda_q|$ remain the same. The dual
quarks will be in the DC - phase with the constituent mass ${\ov \mu}_C=|\langle{\ov q}q
\rangle |^{1/2}=(m_Q\cm)^{1/2}\ll |\Lambda_q|$ if ${\ov \mu}_C\gg m^{\rm pole}_q$\,, where
$m^{\rm pole}_q$ is the pole mass of dual quarks. This is fulfilled, see (8.1)\,:
\bq
\frac{m^{\rm pole}_q}{|\Lambda_q|}=\Bigl (\frac{m_q}{|\Lambda_q|} \Bigr )^
{\frac{1}{1+\gamma_q}}=\Bigl (\frac{{\ov \mu}_C}{|\Lambda_q|} \Bigr )^2\ll
\frac{{\ov \mu}_C}{|\Lambda_q|}\ll 1\,,\quad \lym\ll{\ov \mu}_C\ll |\Lambda_q|\,.\nonumber
\eq

So, the Lagrangian of mions and nions will have the form (5.5),(5.6), with the only replacement $\la
\ra\cm$ in the mion Kahler term and in the first term of the superpotential. So, instead of (8.2), 
the masses of mions and nions (with the logarithmic accuracy) are now \,:
\bq
\frac{\mu_M}{|\Lambda_q|}\sim \frac{\mu_N}{|\Lambda_q|}\sim \Biggl (\frac{{\ov \mu}^2_C}{z_M
|\Lambda_q|^2}\Biggr )^{1/2}=\Biggl(\frac{{\ov \mu}_C}{|\Lambda_q|}\Biggr )^{(N_F+N_c)/2N_c}\,,
\quad z_M=\Bigl (\frac{{\ov \mu}_C}{|\Lambda_q|} \Bigr )^{\gamma_M}\gg 1\,.
\eq

\vspace{0.5cm}
Let us finish this section with a short discussion of a possible behavior of the direct theory
in the case $m_Q\gg \la$. We have to start then from the UV-region $\mu=M_{o}$, supposing that
this theory is considered as the effective low energy theory with the UV cutoff $M_{o}$.

Let us use (2.7) for $N_F>3N_c$. It is seen that the hierarchy of the standard scale parameters
at $\mu=\la$ and $N_F>3N_c\,,\,\,m_Q\gg \la$ remains the same as it was at $N_F<3N_c$ and
$m_Q\ll \la$, i.e.\,: $\cm \gg \Lambda_{YM}\gg m_Q$. But what is really the highest physical
scale $\mu_{H}$ depends on a competition between $\cm$ and the quark pole mass $m_Q^{\rm pole}$.
The value of this last depends on the value of the quark anomalous dimension $\gamma_{Q}$. If
$\cm>m_Q^{\rm pole}$\,, the theory will be in the DC - phase, while
at $m_Q^{\rm pole}>\cm$ it will be in the $\rm HQ$ (heavy quark) phase.

For definiteness, let us use the same value of $\gamma_{Q}$ as in (7.2) with $\gamma_q\ra 0$\,:
\bq
\gamma_{Q}=(2N_c-N_F)/(N_F-N_c)<0\,\quad {\rm at}\quad  N_F>3N_c\,.
\eq
Then
\bq
\frac{m_Q^{\rm pole}}{\la}=\Biggl (\frac{m_Q}{\la} \Biggr )^{\frac{1}{1+\gamma_Q}}=
\Biggl (\frac{m_Q}{\la} \Biggr )^{\frac{N_F-N_c}{N_c}}\gg \frac{\cm}{\la}=
\Biggl (\frac{m_Q}{\la} \Biggr )^{\frac{N_F-N_c}{2N_c}}\,,\quad \frac{m_Q}{\la}\gg 1\,.
\eq
Therefore, with this value of $\gamma_Q$, when going from high UV $\mu=M_o \gg m_Q^{\rm pole}$
down to lower energies, the highest physical scale encountered is $\mu_H=m_Q^{\rm pole}$. The
quarks will be in the $\rm HQ$ (heavy quark) phase.

After integrating out all quarks as heavy ones, one remains with the pure Yang-Mills theory,
but now {\it in the strong coupling regime,\, $a_{-}=N_c\alpha(\mu=m_Q^{\rm pole})/2\pi\gg 1$}.
So, the matching of couplings at $\mu=m_Q^{\rm pole}$ looks now as follows. The coupling of
the higher energy theory is, see (7.4)\,:
\bq
a_{+}=\Biggl (\frac{m_Q^{\rm pole}}{\la} \Biggr )^{-\nu\,\,=\Bigl (\frac{2N_F-3N_c}{N_F-N_c}
\Bigr )}=\Biggl (\frac{m_Q}{\la} \Biggr )^{\frac{2N_F-3N_c}{N_c}}\gg 1\,.
\eq
It follows from the perturbative NSVZ $\beta$ - function \cite{NSVZ} that the coupling of the
lower energy Yang-Mills theory in the strong coupling regime is\,: $a_{-}(\mu\gg \lambda_{YM})
=(\mu/\lambda_{YM})^3$\,. So,
\bq
a_{-}=\Biggl (\frac{m_Q^{\rm pole}}{\lambda_{YM}} \Biggr )^{3}=a_{+}\,\,\, \ra \,\,\,
\lambda_{YM}=\Biggl (\la^{\bo}\det m_Q \Biggr)^{1/3N_c}=\lym\gg \la\,.
\eq

We have now the Yang-Mills theory in the strong coupling perturbative regime at $\lym\ll\mu<
m_Q^{\rm pole}$\,, with its coupling {\it decreasing} with $\mu$ as $a(\mu)=(\mu/\lym)^3$
until it becomes $O(1)$ at $\mu\sim \lym$\,, and here the non-perturbative effects come into a
game. So, integrating at $\mu < \lym$ all gauge degrees of freedom, except for the one whole
field $S\sim W^{2}_{\alpha}$\,, and using the VY - form for the superpotential of $S$ \cite{VY}
one obtains the right value of the gluino condensate, $\langle S\rangle=\lym^3$\, (and a large
number of gluonia with the mass scale $\sim \lym$\,).

On the whole, the mass spectrum includes only two mass scales in this case\,: a large number
of heavy flavored quarkonia with the mass scale $\sim m_Q^{\rm pole}\gg \lym$, and a large
number of gluonia with the universal mass scale $\sim\lym\gg \la$.\\

\section { Conclusions}

\hspace {6 mm} As was described above, within the dynamical scenario considered in this paper,
the direct SQCD theory is in the DC (diquark-condensate) phase at $N_c<N_F<3N_c$. In this case, its
properties and the mass spectrum were described and compared with those of the dual theory.
 It was shown that the direct and dual theories  are different, in general. The only
region where no difference was found up to now, is the case when both theories are in
the perturbative superconformal regime (see above). All this can be of significance in a
wider aspect\,,\,-\, as a hint that many of various dualities considered in the literature
can be strictly valid, at best, also in the superconformal regime only.

We will not repeat here in detail the above described results. Rather, let us compare the
gross features of SQCD and ordinary QCD.
The above described properties of SQCD at $N_c < N_F < 3N_c$ resemble, in many respects,
those of QCD.
\footnote{\,
QCD  means here our QCD with $N_c=3$ and $N_F\simeq 3$ of light flavors.
}
I.e., there is simultaneously confinement and chiral flavor
symmetry breaking, with formation of heavy constituent quarks and light pions.
Besides, in both theories there is a large number of (quasi) stable heavy quarkonia and
gluonia. The main difference is in the parametrical dependence of different observable
masses in the spectrum on the fundamental parameters of Lagrangians: $\la$ and the current
quark masses $m_Q=m_Q(\mu=\la)$, when  $m_Q\ll\la$.\\

a) The scale of the chiral symmetry breaking $\Lambda_{\rm ch}$ (and so the masses of
constituent quarks) is $\Lambda_{\rm ch}^{QCD}\sim \la$ in QCD, while it is parametrically
smaller in SQCD: $\Lambda_{\rm ch}^{SQCD}\sim \cm=(\la^{\bo}m_Q^
{\nd})^{1/2N_c}\ll \la$.

b) The confinement scale (i.e. the string tension $\sqrt \sigma$\,) is $\Lambda_{\rm conf}^{QCD}
=(\sigma_{QCD})^{1/2}\sim \la \sim \Lambda_{\rm ch}^{QCD}$ in QCD, while it is parametrically
smaller than even $\Lambda_{\rm ch}^{SQCD}$ in SQCD: $\Lambda_{\rm conf}^{SQCD}=(\sigma_{SQCD})^
{1/2}\sim \Lambda_{YM}=(\la^{\bo}m_Q^{N_F})^{1/3N_c} \ll\Lambda_{\rm ch}^{SQCD}\sim\cm \ll \la$.

c) So, the heavy quarkonia (meson and baryon) masses are also parametrically different:
$M_{\rm meson}^{QCD}\sim (\Lambda_{\rm ch}^{QCD}+\Lambda_{\rm conf}^{QCD})\sim \la,\, M_{\rm baryon}
^{QCD}\sim N_c \Lambda_{\rm ch}^{QCD}\sim N_c\la$ in QCD, while they are $M_{\rm meson}^
{SQCD}\sim \cm \ll \la,\,\,M_{\rm baryon}^{SQCD}\sim N_c\,\cm$ in SQCD.

d) The masses of gluonia are $M_{\rm gl}^{QCD}\sim \Lambda_{\rm conf}^{QCD}\sim \la$ in QCD,
while they are $M_{\rm gl}^{SQCD}\sim \Lambda_{\rm conf}^{SQCD}\sim \Lambda_{YM}\ll \cm \ll
 \la$ in SQCD.

e) The smallest pion masses are $M_{\pi}^{QCD}\sim (m_Q\Lambda_{\rm ch}^{QCD})^{1/2}\sim
(m_Q\la)^{1/2}\gg m_Q$ in QCD, while they are not $\sim (m_Q\cm)^{1/2}$,\, but
$M_{\pi}^{SQCD}\sim m_Q$ in SQCD (this last
difference is because the spin $1/2$ quarks are condensed in QCD, while these are spin
zero quarks in SQCD).\\

Now, let us comment briefly on the $N_c$\,-\,dependence of various quantities that
appeared in the text above. The standard $N_c$\,-\,counting rules predict that
the gluino and quark condensates, $\langle S\rangle$ and $\langle{\ov Q}^{\ov j}Q_i\rangle$,
are not $O(1)$ at $N_c\gg 1\,,\,N_F/N_c={\rm const}$\,, as in the text, but $N_c$ times
larger, $O(N_c)$ (and this agrees with explicit calculations, see e.g. \cite{SV-r}).
\footnote{\,
Besides, this can be seen from the example with $N_F<N_c$\,, when quarks are higgsed (see
section 2). The gluon masses, $\mu_{\rm gl}^2\sim \alpha(\mu=\mu_{\rm gl})\,{\cal M}^2_{o}$\,,
are $O(1)$. Because $\alpha=O(1/N_c)\,,\,{\cal M}^2_{o}$ is $O(N_c)\,, \langle S\rangle=
{\hat m}_Q\,{\cal M}^2_{o}=O(N_c)\,.$

Connected with this, there is the inherent ambiguity in the VY\,-\,procedure for the pure
Yang-Mills theory: one can replace $\ln (\mu^3/\Lambda^3)$  by $\ln (S/C_o\Lambda^3)-1$,
where $C_o$ is some constant. The value $C_o=1$ was used everywhere in the text, while
$\mu^3$ is definitely $N_c$\,-\,independent, so that a better replacement is rather\,:
$\ln(\mu^3/\Lambda^3)\ra \ln (S/N_c\Lambda^3)-1$\,, resulting in $\langle S \rangle=N_c\,
\Lambda^3$.
}
The right dependence on $N_c$ can easily be restored over all the text by simple
substitutions, for instance, $\la^{\bo}\ra N_c^{N_c}\la^{\bo}$ in (3.13), etc.\\

Finally, we would like to make a comment about the spontaneously SUSY-breaking
metastable local vacuum in SQCD with $N_c+1<N_F<3N_c/2,\,\,m_Q\neq 0,\, m_Q\ll
\la$, proposed recently in \cite{ISS}. The arguments for the existence of such
a state in the dual theory are presented in \cite{ISS}.

Recalling general arguments given above in section 7 (\,see (7.2)-(7.4), it is worth
also recalling that these arguments are not connected with the use of the dynamical
scenario with the diquark-condensate) that the direct and dual theories are not
equivalent in the infrared region, it becomes insufficient to show such a state in the
dual theory, because this does not imply automatically that this state exists also
in the direct theory. So, let us try to find out this state within the direct theory.

In terms of the direct theory fields, this state is characterized  by all $N^2_F$ components
$\langle M^i_{\ov j}\rangle=\langle {\ov Q}_{\ov j}Q^i \rangle=0\,,$\, while $\langle B\rangle
={\rm const}\langle b\rangle\neq 0\,$ (and $\langle {\ov B}\rangle$ the same),\, $B\ra Q^{N_c},
\, b\ra q^{\nd}$. Unfortunately, no simple possibility for a local vacuum with these properties
is seen in the direct theory.  For instance, the dynamics underlying the appearance of the
above basic nonzero baryon condensates looks obscure. If these baryon condensates were,
for instance, due to higgsed quarks $\langle Q^{i}\rangle=\langle{\ov Q}_{\ov j}\rangle\neq 0$\,,
with $i\,,{\ov j}=1...N_c$\,, so that $\langle B \rangle=\langle {\ov B} \rangle\sim \langle
Q^{i}\rangle ^{N_c}\neq 0$, then no reason is seen for all components of $\langle M^i_{\ov j}
\rangle=\langle {\ov Q}_{\ov j}Q^i \rangle$ to be exactly zero. Rather, $\langle M^i_{\ov j}
\rangle$ with $i={\ov j}=1...N_c$ will be $\sim \langle Q^{i}\rangle\langle{\ov Q}_{i}\rangle
\neq 0$\,. Besides, looking at the Lagrangian in (3.2) it is seen that it
becomes singular at $\cm\ra 0$. So, it seems impossible that the local vacuum
with the above given properties can appear here.

However, this is not the whole story as (3.2) is a local Lagrangian, i.e. it is valid only
locally in the field space, not too far from the genuine SUSY-vacuum. This implies that
generally, besides $M^i_{\ov j}$\,, the additional fields can be involved to describe
correctly the vicinity of the above metastable vacuum. So, let us try in addition from
another side, using some specific properties of the above metastable state of the dual
theory. Let us also look at the lightest excitations around this vacuum. As was argued in
\cite{ISS}, all excitations have masses $\sim (m_Q\la)^{1/2}$, except for some massless
modes of the baryon and $M^i_{\ov j}=({\ov Q}_{\ov j}Q^i)$ fields (and the basic vacuum
condensates of baryons). So, let us take the scale $\mu \ll (m_Q\la)^{1/2}$ and try to
write by hand the effective superpotential made of these meson and baryon fields
only. The simplest form is:
\bq
W_{\rm eff}=-\nd {\rm \Biggl \{\frac{ \det M-Tr\,({\ov B}\,M^{\nd} B)}{\la^{\bo}}
\Biggr \}^{1/\nd}+m_Q Tr\, M\,}.
\eq

At $\nd\geq 2$ no possibility is seen to obtain from (9.1) the non-singular expansion in
quantum fluctuations around the state with $\langle M \rangle=0\,,\, \langle B\rangle=\langle
{\ov B}\rangle \neq 0$\,.
\footnote{\,
Formally, one can multiply the first term in the r.h.s. of (9.1) by a function $f(z)\,,
\,\,z={\rm \det M/Tr\,({\ov B}\,M^{\nd} B)}$\,, but this does not help to avoid singularities.
}
Only the case $\nd=1$ is non-singular in (9.1). But even in this case one has to show
then how it is possible to obtain in some way (9.1) starting with (2.1) and expanding
self-consistently around this metastable vacuum. It seems, there will be problems.

Finally, the absence of the above metastable spontaneously SUSY-breaking state
in the direct theory may be not so surprising, taking into account all arguments
given above in the text that the direct and dual theories are not equivalent.\\

This work is supported in part by the RFBR grant 07-02-00361-a.\\

\hspace {2cm} {\Large \bf Appendix}\\

\hspace {6 mm} The purpose of this appendix is to comment in short on a situation with anomalous
divergences of external currents
(the 't Hooft triangles) in SQCD, within the dynamical scenario considered in this paper.\\

In our ordinary QCD, at the scale $\mu_{\rm ch}\sim \la$ and at $m_Q\ra 0$\,, there is the genuine
spontaneous breaking of the flavor symmetry: $SU(N_F)_L\times SU(N_F)_R \ra SU(N_F)_{L+R}$\,, while the
baryon symmetry $U(1)_B$ remains unbroken. So, the quarks acquire the constituent masses $\mu_C\sim
\mu_{\rm ch}$ and decouple at $\mu<\mu_{\rm ch}$ (together with all gluons which acquire (either
electric or magnetic) masses $\sim \la$ due to non-perturbative confining interactions, so that
the lower energy theory contains only $(N_F^2-1)$ light pions. If the quarks are exactly massless,
the pions are also massless, while if the chiral
symmetry $SU(N_F)_L\times SU(N_F)_R$ is broken explicitly down to $SU(N_F)_{L+R}$ by parametrically
small quark masses $0<m_Q\ll \la$\,, the pions become the pseudo-Goldstone bosons with
parametrically small masses $m_{\pi}\sim (m_Q\la)^{1/2}\ll \mu_{\rm ch}$\,.

In SQCD with $N_F<N_c$ and with small {\it explicit} breaking of chiral flavor symmetry and R-charge
by quark masses $0<m_Q\ll \la$ (see section 2), the scalar quarks are higgsed at the high scale
$\mu_{\rm ch}=\mu_{\rm gl}\gg \la \,\, (\mu_{\rm gl}\simeq \cm$\,, with the logarithmic accuracy) and
acquire the large "constituent masses" $\mu_C=\mu_{\rm gl}$. The color symmetry $SU(N_c)$ is broken
down to $SU(N_c-N_F)$\,, and $(2N_c N_F-N_F^2)$ gluons become massive eating the Goldstone bosons. So,
all this can be considered as the quasi-spontaneous symmetry breaking: $SU(N_c)_{C}\times SU(N_F)_L
\times SU(N_F)_R\times U(1)_R\times U(1)_B\ra SU(N_c-N_F)_C\times SU(N_F)_{C+L+R}\times U(1)_B$\,,
as the "constituent masses" $\mu_C\sim \cm$ are parametrically larger than the pion masses $m_{\pi}
\sim m_Q$ (with the logarithmic accuracy). As a result, there appear $N_F^2$ pseudo-Goldstone pions
(together with their superpartners). So, the lower energy theory at $\mu<\mu_{
\rm gl}$ includes the superfields of light $(N_c-N_F)^2-1$ gluons and $N_F^2$ pions.

In SQCD with $N_F>N_c$ and $m_Q\ll \la$ (in the dynamical scenario considered in this paper), all
quarks acquire the constituent masses $\mu_C=\cm\ll \la$ in the threshold region $\mu\sim\mu_{\rm ch}=
\cm$\,, and there appear $N_F^2$ light pions, while all gluons remain massless. This also can be
considered as the quasi-spontaneous symmetry breaking: $SU(N_F)_L\times SU(N_F)_R\times U(1)_R\times
U(1)_B\ra SU(N_F)_{L+R}\times U(1)_B$\,, as the constituent quark masses $\mu_C$ are parametrically
larger than the pion masses $m_{\pi}\sim m_Q\ll \cm$. The lower energy theory at $\mu<\cm$
includes the superfields of light $(N_c^2-1)$ gluons and $N_F^2$ pions.\\

Let us recall now some important and well known properties of the lower energy theory at $\mu<\mu_
{\rm ch}$.\\
1) After integrating out all heavy fields (and all Fourier components of light fields with
$k>\mu_{\rm ch}$)\,, the Lagrangian of the lower energy theory at $\mu<\mu_{\rm
ch}$ will be {\it local}, right because all integrated modes were hard (it is always implied that
this integration is performed in a way which respects all symmetries).\\
2) The external global symmetries can be gauged by introducing external vector fields and adding the
appropriate set of massless "leptons", so that all anomalous divergences of external currents
originating from the quark-gluon sector will be canceled by those originating from the lepton one.\\
3) After all this, because the symmetry breaking in the quark-gluon sector was quasi-spontaneous,
the lower energy Lagrangian {\it will preserve all previous symmetries, both internal and external}.
So, because nothing happens with leptons when crossing the scale $\mu=\mu_{\rm ch}$\,, the anomalous
divergences originating from the quark-gluon sector also remain the same \cite {hooft}.

So, there is no questions whether the lower energy theory behaves properly under symmetry
transformations, both internal and external, or whether the anomalous divergences of external currents
originating from the quark-gluon sector will remain the same in the lower energy theory,
\footnote{\,
i.e. at scales $\mu_{\rm expl}<\mu<\mu_{\rm ch}$, where $\mu_{\rm expl}\sim m_{\pi}\ll \mu_{\rm ch}$
is the scale of the {\it explicit} global chiral symmetry breaking, because the explicitly broken
global symmetry is incompatible with gauging this symmetry, and $\mu_{\rm expl}$
can be neglected only at scales $\mu>\mu_{\rm expl}$\,.
}
as they were in the higher energy theory at $\mu>\mu_{\rm ch}$, - {\it this is automatic}. The only
relevant questions are\,: a) what is the explicit form of the lower energy Lagrangian; b) in what
way, explicitly, the anomalous divergences of external currents originating from the quark-gluon
sector are saturated by fields of the lower energy theory.

As for 'a', if the dynamics of the theory is under a full control, the explicit form of the lower
energy Lagrangian is obtained by the above described direct integration. As is well known, besides
the 'standard terms', there will appear additional Wess-Zumino-like terms \cite{WZ}\cite{GW}.

Now, a few words about 'b', within the dynamical scenario for SQCD considered in this paper. First, as
for pions, it is worth noting that because contributions of pion loops are power suppressed
at scales $\mu<\mu_{\rm ch}$, these loops will give only small power corrections to the contributions
of tree diagrams into amplitudes with low energy external pions and/or external gauge fields.

There will appear one-pion terms, $J^{\rm ext}_{\nu}\sim iF_{\pi}\partial_{\nu}\pi+\dots$\,, in those
external currents which correspond to quasi-spontaneously broken generators, with the pion decay
constant $F_{\pi}\sim {\cal M}_o$ for $N_F<N_c$\,, and $F_{\pi}\sim \cm$ for $N_c<N_F<3N_c$. Besides,
among many others, there will be the well known term $\sim F_{\pi}^{-1}{\rm Tr}(\pi F_{\mu\nu}{\tilde F}
_{\mu\nu})$ in the Wess-Zumino part of the Lagrangian (here $F_{\mu\nu}$ is the field strength of the
external vector fields, $W_{L}$ or $W_{R}$ - bosons, or R-photon $A^{R}$), with the appropriate
coefficient. As a result, the anomalous divergences of all such currents will be automatically
saturated by a sum of three contributions: a) the one intermediate pion exchange, b) the direct
contributions into triangles of fermionic pion superpartners, c) the additional direct contributions
of gluinos into $R$ and $R^3$ triangles.

So, for instance, for all $N_c<N_F<3N_c$ (with the logarithmic accuracy for $N_F<N_c$), the decay
width of the pion into two (sufficiently light at small $\alpha_{\rm ext}$) vector bosons $V=\{W_L,\,
W_R,\,A^{R}\}$  will be: $\Gamma(\pi\ra 2V)\sim \alpha_{\rm ext}^2m_{\pi}^3/F^2_{\pi}\sim\alpha_
{\rm ext}^2 m_Q^3/\cm^2\sim \alpha_{\rm ext}^2\la(m_Q/\la)^{\Delta},\, \Delta={(4N_c-N_F)/N_c}$.

Those external currents, e.g. the baryon one, which corresponds to the unbroken generators, will not
contain the one-pion term (because there is no corresponding pion), and their
anomalous divergences, like $\langle W_L|\partial_{\nu}J_{\nu}^{B}|W_L\rangle$\,, will be directly
saturated by the point-like terms $\sim(\epsilon_{\nu\lambda\sigma\tau} A_{\nu}^{B}W^{L}_{\lambda}
\partial_{\sigma}W^{L}_{\tau}+\dots)$ in the Wess-Zumino part of the Lagrangian.\\

We did not write explicitly in the main text the Wess-Zumino-like terms because: a) this is not a
simple matter to find their explicit form, b) they are irrelevant for the main purpose of this
paper - to calculate the mass spectrum of the theory.\\

\end{document}